\documentclass[useAMS,usenatbib,fleqn]{mn2e}

\usepackage{graphicx}
\usepackage{float}
\usepackage{epstopdf}
\usepackage{amsmath}
\usepackage{amssymb}
\usepackage{epsfig}
\usepackage{lscape}
\usepackage{rotating}
\usepackage{subfigure}
\usepackage{color}

\def\gtsim {\gtrsim}   
\def\ltsim {\lesssim}   

\newcommand{\dummytitle}[1]{}
\newcommand{\msun}{{\rm M}_\odot}
\newcommand{\ltgt}{LTG$_{\Delta {\rm t}}$ }
\newcommand{\etgt}{ETG$_{\Delta {\rm t}}$ }
\newcommand{\ltgs}{LTG$_{\Delta {\rm s}}$ }
\newcommand{\etgs}{ETG$_{\Delta {\rm s}}$ }

\defcitealias{pt14}{TK14}
\defcitealias{pt15a}{TK15a}
\defcitealias{pt15b}{TK15b}

\title[Metallicity Gradients in Simulated Galaxies]{The Metallicity and Elemental Abundance Gradients of Simulated Galaxies, and their Environmental Dependence}
\author[P.~Taylor and C.~Kobayashi]{Philip~Taylor\thanks{E-mail: philip.1.taylor@anu.edu.au}$^1$ and Chiaki~Kobayashi$^2$\\
$^1$Research School of Astronomy and Astrophysics, The Australian National University, Canberra, ACT 2611, Australia\\
$^2$Centre for Astrophysics Research, School of Physics, Astronomy and Mathematics, University of Hertfordshire, Hertfordshire, AL10 9AB, UK}

\begin{document}

\date{Accepted  Received ; in original form}

\pagerange{\pageref{firstpage}--\pageref{lastpage}} \pubyear{}

\maketitle

\label{firstpage}

\begin{abstract}
The internal distribution of heavy elements, in particular the radial metallicity gradient, offers insight into the merging history of galaxies.
Using our cosmological, chemodynamical simulations that include both detailed chemical enrichment and feedback from active galactic nuclei (AGN), we find that stellar metallicity gradients in the most massive galaxies ($\gtsim 3\times10^{10}\msun$) are made flatter by mergers and are unable to regenerate due to the quenching of star formation by AGN feedback.
 {The fitting range is chosen on a galaxy-by-galaxy basis in order to mask satellite galaxies.}
The evolutionary paths of the gradients can be summarised as follows; i) creation of initial steep gradients by gas-rich assembly, ii) passive evolution by star formation and/or stellar accretion at outskirts, iii) sudden flattening by mergers.
There is a significant scatter in gradients at a given mass, which originates from  {the last path, and therefore from galaxy type.
Some variation remains at given galaxy mass and type because of the complexity of merging events, and hence we find only a weak environmental dependence.}
Our early-type galaxies (ETGs), defined from the star formation main sequence  {rather than their morphology}, are in excellent agreement with the observed stellar metallicity gradients of ETGs in the SAURON and ATLAS$^{\rm 3D}$ surveys.
We find small positive [O/Fe] gradients of stars in our simulated galaxies, although they are smaller with AGN feedback.
Gas-phase metallicity and [O/Fe] gradients also show variation, the origin of which is not as clear as for stellar populations.
\end{abstract}



\section{Introduction}
\label{sec:intro}
To understand the formation and evolution of galaxies, two series of observations should be consistently explained.
One is the scaling relations of galaxies, whereby global physical properties (e.g., size, metallicity) of galaxies correlate with galaxy mass, which we focused on in our previous paper \citep[][hereafter \citetalias{pt15a}]{pt15a}.
The other observation is the internal structure of galaxies, namely metallicity radial gradients of stars and gas within galaxies.
Chemical elements are fossils in galactic archaeology, on which the star formation and chemical enrichment histories are imprinted \citep[e.g.,][]{nomoto13}.
Therefore, metallicity gradients are supposed to give one of the most stringent constraints; for disc galaxies, the radial gradients suggest the inside-out growth \citep[e.g.,][]{ck11a}, and they can also constrain the merging histories of the galaxies \citep[e.g.,][]{ck04}.

Major mergers are one of the great drivers of morphological change \citep[e.g.,][]{toomre72,barnes88}, and can have profound effects on the subsequent evolution of a galaxy.
Recent ($\ltsim 1$ Gyr) major mergers can be observationally identified from dynamical disturbances, including tidal structures, shells, and complex internal kinematics \citep[e.g.,][]{schweizer92}.
On the other hand, metallicity gradients could also trace major merges in the past.
However, it is not straightforward to constrain the evolutionary history of a galaxy from only its gradients because of the complexity of the evolution due to various physical processes.
The following evolutionary paths have been suggested: i) monolithic collapse forms steep initial gradient \citep[e.g.,][]{larson74,carlberg84,pipino10}; ii) gradient passively evolves, becoming flatter; iii) gradient suddenly flattened by major merger \citep[e.g.,][]{white80}; iv) following a gas-rich (`wet') merger, central star formation can regenerate gradient \citep[e.g.,][]{hopkins09}.
\citet{ck04} studied how mergers alter the stellar metallicity gradients of ETGs in a cosmological context, and found that major mergers indeed cause flattening, with mergers of larger mass ratios exerting the greatest influence.
Conversely, mergers with gas-rich galaxies induced central star formation, which caused steepening of the gradients.
With the combination of these effects, significant variation of gradients is found at a given mass.
In fact, the lack of tight correlation between gradients and mass has been found and discussed with slit observations of a limited sample of ETGs \citep[e.g.,][]{davies93,ck99,ogando05,spolaor10}.

In this paper, we present both stellar and gas-phase metallicity gradients of the galaxies in our cosmological simulations, where various types and masses of galaxies are included based on $\Lambda$CDM cosmology.
Therefore, our prediction can be directly compared with the ongoing and future observational surveys with integral field spectrographs such as SAURON \citep[e.g.,][]{kuntschner10}, CALIFA \citep[e.g.,][]{califa12}, SAMI \citep[e.g.,][]{ho14} and its successor HECTOR \citep[e.g.,][]{hector15}, and MaNGA \citep[e.g.,][]{bundy15}.
These surveys are designed as the internal structure is supposed to be the key to understand galaxy formation and evolution, and the metallicity gradients will also be obtained for thousands of nearby galaxies over the coming decades.

In order to predict metallicity gradients, it is necessary to use hydrodynamical simulations that include star formation, feedback, and detailed chemical enrichment.
It is also necessary to quench star formation at certain epochs depending on the galaxy mass as shown in observations, i.e. the so-called downsizing effect \citep[e.g.][]{cowie96} where the quenching occurs earlier in more massive galaxies.
Theoretically, the feedback from active galactic nuclei (AGN) is the most popular source \citep[e.g.,][]{croton06}, and has also been included in our hydrodynamical simulations.
Compared with other hydrodynamical simulations, \citep[e.g.,][]{springel05,schaye15,sijacki15}, our AGN feedback scheme is unique in terms of seeding black holes (BH), which is motivated from the first star formation \citep{pt14}.
Our AGN model gives good agreement with many observations such as cosmic star formation rates, BH mass--bulge mass relation, galaxy size--mass relation, and mass--metallicity relations \citepalias{pt15a}.
As shown in \citet{pt15b}, our AGN feedback causes  large-scale, metal-enhanced outflows, which could impact the gas-phase metallicity gradients of host galaxies.

This paper is arranged as follows: in Section \ref{sec:5paper5_fit} we give a brief summary of our cosmological simulations and describe the procedure used to measure the metallicity gradients of simulated galaxies, as well as introducing parameters derived from our simulation that correlate with galaxy morphology.
The present-day distribution of stellar gradients of $Z$ and [O/Fe] is analysed in Section \ref{sec:5paper5_starz0}.
In Section \ref{sec:5paper5_res_gradt} we describe the evolution of the stellar metallicity gradients of two example galaxies \citepalias[A and B of][]{pt15a}, which have different masses and evolve in different environments, across cosmic time.
In Section \ref{sec:5paper5_massmet} we present gas-phase metallicity gradients at $z=0$, as they give another constraint on the enrichment history of the galaxies.
Finally, we give our conclusions in Section \ref{sec:5paper5_conc}.

\section{Simulation Setup and Analysis Procedure}
\label{sec:5paper5_fit}

\subsection{Our Simulations}

Our simulations use a chemodynamical simulation code \citep{ck07,pt14}, which is based on the {\sc gadget-3} code \citep{springel05}, updated with physical processes relevant to galaxy formation and evolution.
We include a prescription for star formation \citep{ck07} assuming a \citet{kroupa08} IMF, as well as energy feedback and chemical enrichment from AGB stars \citep{ck11b}, Type Ia and Type II supernovae \citep[SNe,][]{ck04,ck09}, and hypernovae \citep{ck09,ck11a}.
BH physics is also included: BHs form from high density, metal-free gas; they grow via Eddington-limited, Bondi Hoyle accretion, and mergers with other BHs; energy from gas accretion is distributed to neighbouring gas particles in a purely thermal form.
A more detailed description of our BH model can be found in \citet{pt14}.
We also include metallicity-dependent radiative gas cooling \citep{sutherland93}, and heating from an evolving, uniform UV background \citep{haardt96}.

Our simulations are run in a $(25h^{-1}$ Mpc$)^3$ box with periodic boundary conditions; one includes the BH physics described above, the other does not.
We adopt a 9-year \emph{Wilkinson Microwave Anisotropy Probe} $\Lambda$CDM cosmology \citep{wmap9} with $h=0.70$, $\Omega_{\rm m}=0.28$, $\Omega_\Lambda=0.72$, $\Omega_{\rm b}=0.046$, and $\sigma_8=0.82$.
The initial conditions consist of $240^3$ particles of each of gas and dark matter with masses $M_{\rm DM}=7.3\times10^7h^{-1}\msun$ and $M_{\rm gas}=1.4\times10^7h^{-1}\msun$.
We use a gravitational softening length $\epsilon_{\rm gas}=1.125h^{-1}$ kpc.
The large-scale evolution of these simulations was shown in Fig. 1 of \citetalias{pt15a}; our simulations reproduce the observed i) cosmic star formation rate ii) BH mass--stellar velocity dispersion relation iii) size--mass relation of present ETGs and iv) stellar and gas-phase mass-metallicity relations, and are fairly consistent with v) colour--magnitude relations vi) specific star formation rates (SFR), and vii) [$\alpha$/Fe]--stellar velocity dispersion relation of ETGs.

 {Note that our simulated galaxies are calculated in a cosmological box, without the zoom-in technique, because AGN feedback can affect Mpc-scale regions, especially in terms of chemical enrichment \citep{pt15b}.
As for other cosmological simulations, although massive galaxies show discy or spheroidal morphology, the numerical resolution is insufficient to apply the bulge-disc decomposition.
However, in order to investigate the link between the morphology and gradients we introduce two morphology indicators in Section \ref{sec:morph}.
In Sections \ref{sec:5paper5_starz}, we compare our predictions to observations that share the same fitting method as in our simulations.
This paper provides the predictions of our simulated galaxies with certain resolution and baryon physics modelling, as the first step for the comparison to ongoing and future observational surveys with a large sample over a wide mass range.}

\subsection{Fitting Method}\label{sec:fit}
\begin{figure*}
	\centering
	\subfigure{\includegraphics[width=0.48\textwidth,keepaspectratio]{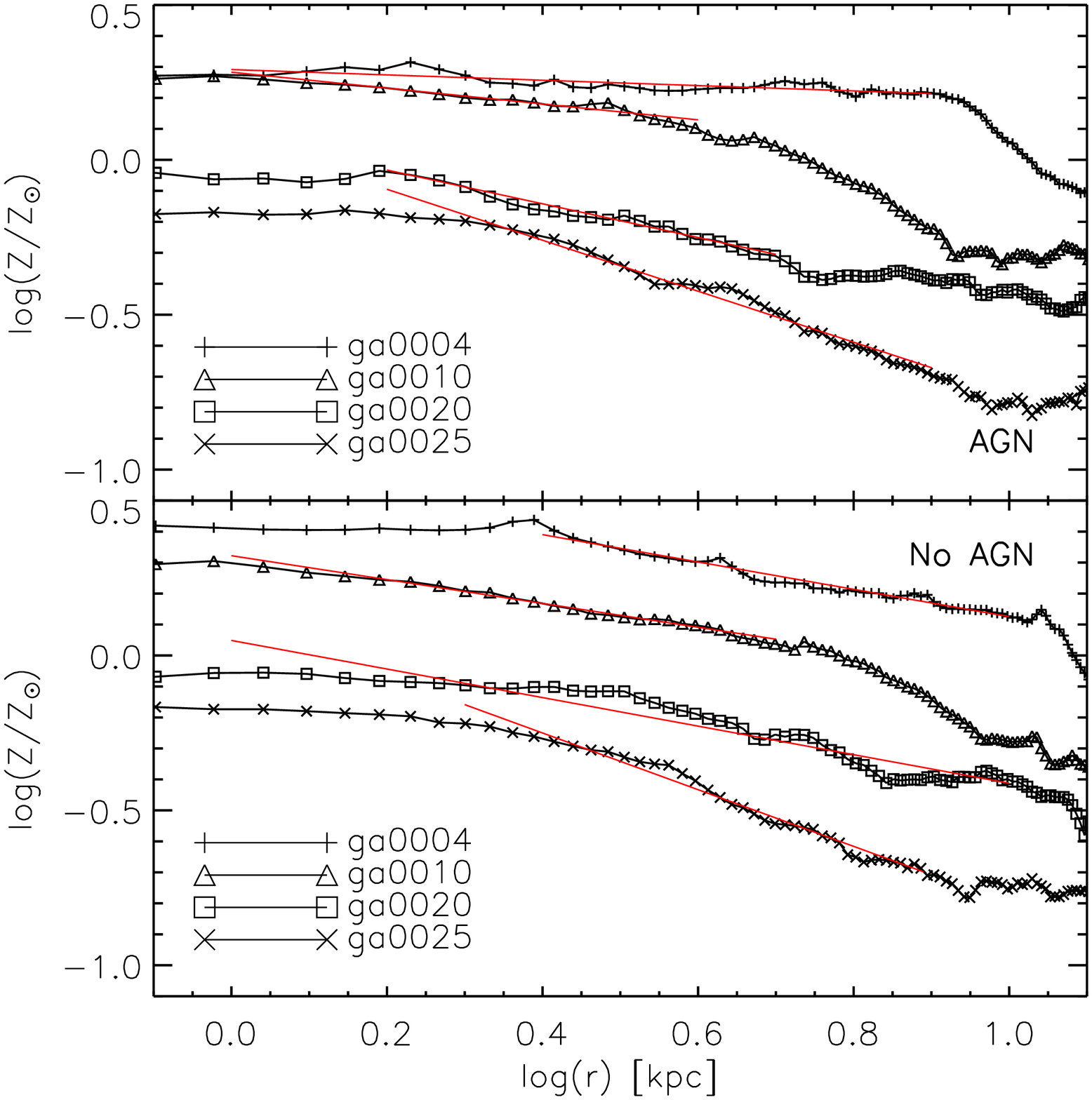}\label{fig:5paper5_fits}}
	\subfigure{\includegraphics[width=0.49\textwidth,keepaspectratio]{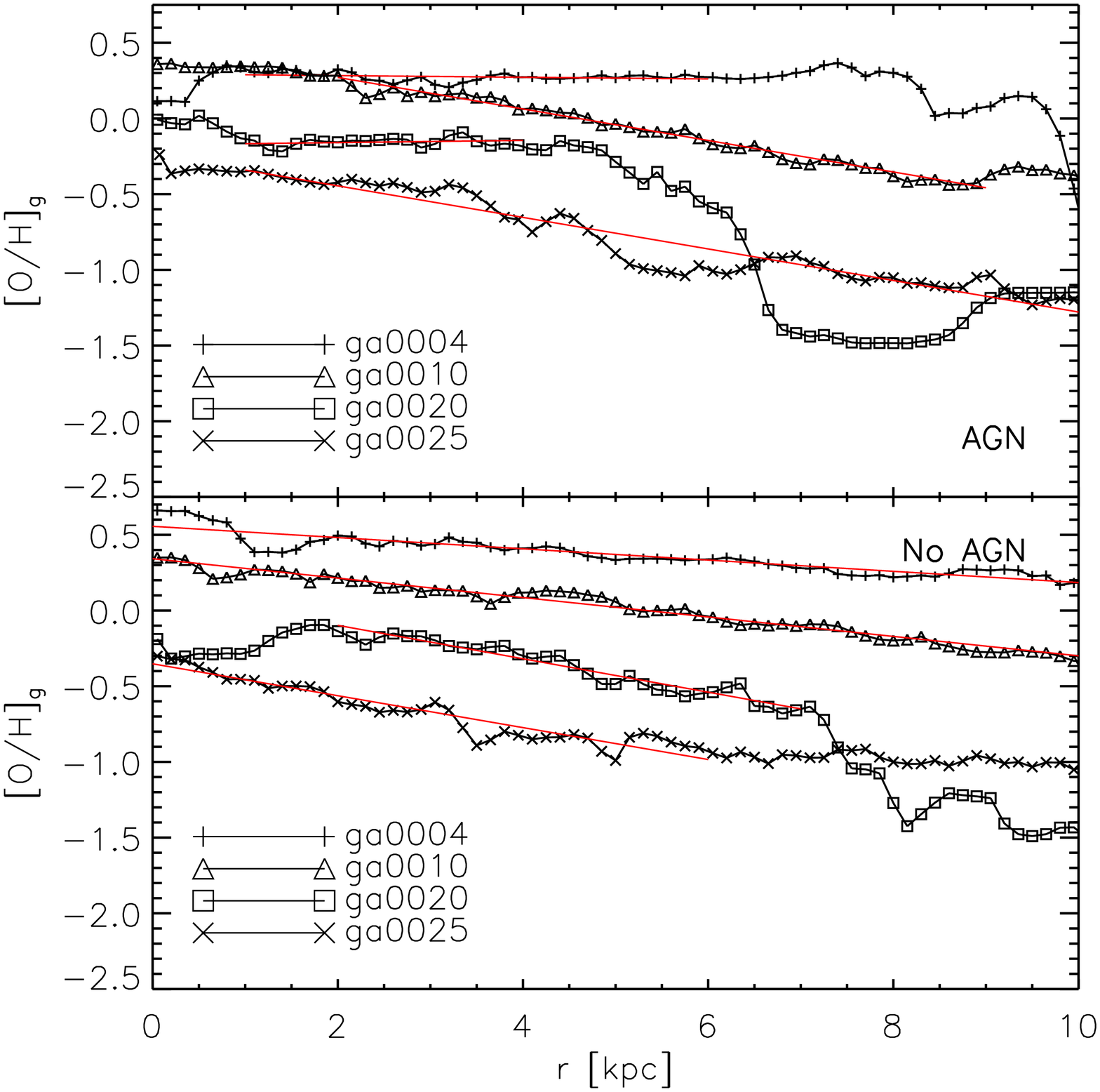}\label{fig:5paper5_fitg}}
	\caption{ {Stellar and gas-phase metallicity profiles (left and right panels, respectively) of four galaxies that are matched between the two simulations and have measurable gradients.
	The top and bottom panels are for the simulation with and without AGN feedback, respectively.
	Solid red lines show the resulting straight-line fit.
	All of the profiles have had a small constant added in the vertical direction for clarity.}}
	\label{fig:5paper5_fit}
\end{figure*}

In order to analyse metallicity gradients in cosmological simulations, we have developed the following new method, and illustrated with the metallicity profiles of four massive galaxies (Fig. \ref{fig:5paper5_fit}).
Galaxies are identified by our parallel Friends of Friends code, which is based on the Friends of Friends code used in \citet{springel01}, and all star or gas particles in the galaxy's vicinity are considered.
 {Each galaxy is viewed such that the net angular momentum vector of its stars lies along the $z$ axis.
Disc galaxies are therefore viewed face-on.}
Annuli of width $\Delta r = 0.2$ kpc centred on the galaxy's centre of mass are used for the metallicity profile, and each particle should contribute to multiple bins due to the nature of SPH.
The fraction of particle $i$ that contributes to bin $j$ is given by
\begin{equation}\label{eq:integ}
	f_{ij} = \int_{-\infty}^\infty {\rm d}z \int_0^{2\pi} {\rm d}\phi \int_{j\Delta r}^{(j+1)\Delta r} W({\bf r-p}_i ,h) r{\rm d}r,
\end{equation}
where $W(r,h)$ is the smoothing kernel with characteristic length $h$, ${\bf p}_i$ is the position of particle $i$, and ${\bf r} = {\bf 0}$ corresponds to the centre of mass of the galaxy.
We adopt the following form for the smoothing kernel, which is essentially the same as the cubic spline used in the simulations, but is computationally more convenient for obtaining the random numbers of equation \eqref{eq:rdash}.
\begin{equation}
	W({\bf r}) = \left(2\pi (0.3\epsilon)^2\right)^{-3/2} \exp\left(-\frac{1}{2}\frac{{\bf r\cdot r}}{\left(0.3\epsilon\right)^2}\right),
\end{equation}
a Gaussian with $\sigma=0.3\epsilon$, where $\epsilon$ is the softening length of the simulations.
The factor $0.3$ enters in order to closely match the cubic spline kernel\footnote{Matching the spline kernel and Gaussian at the origin gives $\sigma=\left(2^{3/2}\pi^{1/6}\right)^{-1}h\approx 0.2921 h$.}.
We use Monte Carlo integration to evaluate equation \eqref{eq:integ}; for each particle $i$ we generate $N=10^4$ of the following quantities:
\begin{equation}\label{eq:rdash}
	r' = |n|\times 0.3\epsilon;
\end{equation}
\begin{equation}
	\cos \theta' = 2u_{\theta'}-1;
\end{equation}
\begin{equation}
	\phi' = 2\pi u_{\phi'},
\end{equation}
where $n$ is a normally distributed random variable and $u$ a uniformly distributed random variable on the interval $[0,1]$.
$N$ new positions can then be generated from ${\bf p}_i$:
\begin{equation}
	{\bf p} = {\bf p}_i + \left(
	\begin{array}{c}
		r'\sin\theta'\cos\phi'\\
		r'\sin\theta'\sin\phi'\\
		r'\cos\theta'
	\end{array}
	\right).
\end{equation}
Finally, these positions are binned by projected radius in the $x-y$ plane and we estimate $f_{ij}$ as
\begin{equation}
	f_{ij} = N_{ij}/N.
\end{equation}

When estimating the stellar metallicity of each bin, we weight by V-band luminosity, where most of the absorption lines used in observations are located, to facilitate comparison with observational data, and so the metallicity of the $j^{\rm th}$ bin is
\begin{equation}
	Z_j = \frac{\sum_i Z_i L_{{\rm V,}i} f_{ij}}{\sum_i L_{{\rm V,}i} f_{ij}}.
\end{equation}
For calculating O/Fe, this is altered to
\begin{equation}
	({\rm O/Fe})_j = \frac{\sum_i \left({\rm O}_i/{\rm Fe}_i\right) L_{{\rm V,}i} f_{ij}}{\sum_i L_{{\rm V,}i} f_{ij}},
\end{equation}
where ${\rm O}_i \equiv M_{{\rm O,}i}/M_{*,i}$, and similarly for Fe.

For gas-phase metallicities, it is necessary to employ a slightly different approach.
In order to make use of as many particles as possible, we also include star particles with ages $<10^7$ yr, which trace the metallicity and location of the gas they formed from closely.
Then the metallicities of star and gas particles are weighted by the mass of young (OB) stars as follows:
\begin{equation}
	Z_j = \frac{\sum_{i_*} {\rm O}_{i_*} M_{i_{*,{\rm init}}}f_{{i_*}j} + \sum_{i_{\rm g}} {\rm O}_{i_{\rm g}}f_{{i_{\rm g}}j} \times {\rm SFR}_{i_{\rm g}}\times t_{{\rm SF},i_{\rm g}}}{\sum_{i_*} M_{i_*,{\rm init}}f_{{i_*}j} + \sum_{i_{\rm g}} f_{{i_{\rm g}}j}\times {\rm SFR}_{i_{\rm g}}\times t_{{\rm SF},i_{\rm g}}},
\end{equation}
where $i_{\rm g}$ and $i_*$ index the gas and young star particles, respectively, and $M_{i_{*,{\rm init}}}$ denotes initial stellar mass.
SFR denotes star formation rate of gas particles, recorded during the simulations, and the star formation timescale of gas particles is $t_{\rm SF}\equiv t_{\rm dyn}/c_*\equiv \left(c_*\sqrt{4\pi G \rho}\right)^{-1}$.
We use $c_*=0.02$ \citep{murante10,scannapieco12}.
Note that this definition of the gas-phase metallicity is different to the global metallicities shown in \citetalias{pt15a}.

Using the radial profiles obtained, we fit an equation of the form
\begin{equation}\label{eq:5paper5_fit}
	\log x/x_\odot = af\left(r\right)+b,
\end{equation}
where $x = $ Z, O, or O/Fe, and $f\left(r\right) = \log r$ for stars, and $r$ for gas to compare with observational data \citep{ck04,ck11a}.
The desired gradient is then simply the value of $a$ in equation \eqref{eq:5paper5_fit}.
 {In our simulated galaxies, the stellar and gas-phase metallicity gradients are well fitted with $\log r$ and linear $r$, respectively.
In many observational papers, gas gradients are available only for LTGs and fitted against linear $r$ \citep[e.g.,][]{ho15}, while stellar gradients are best estimated for ETGs and bulges, and fitted against $\log r$ \citep[e.g.,][]{kuntschner10} because of the weaker effect of the age-metallicity degeneracy.
Thus we compare our simulations only to these comparable observations.
This excludes \citet{sanchez14}, who fitted metallicity gradients with linear r in disc-dominated areas of face-on spiral galaxies} {; this is discussed in Appendix \ref{ap:sanchez}.}

 {The range used for fitting is important, and \citet{ck04} used 1 kpc and max($2R_{\rm e}$, 10\,kpc) for inner and outer boundaries \citep[see][for more details]{ck04}.
For our galaxies, this range provides fairly good fitting, but not perfect.
Therefore, we choose the range on a galaxy-by-galaxy basis, by looking at the metallicity profiles of all galaxies with $M_*\gtsim6\times10^9\msun$ in our simulations (corresponding to 1000 star particles), so that the peaks associated with satellite galaxies at large radii can be masked, and poorly resolved or otherwise `bad' profiles (e.g., due to ongoing mergers) can be rejected.
The global results in the following sections are not so different from those fitted with Kobayashi (2004)'s mass-dependent range, although our galaxy-by-galaxy fitting produces smaller scatters.}

\begin{figure}
	\centering
	\includegraphics[width=0.48\textwidth,keepaspectratio]{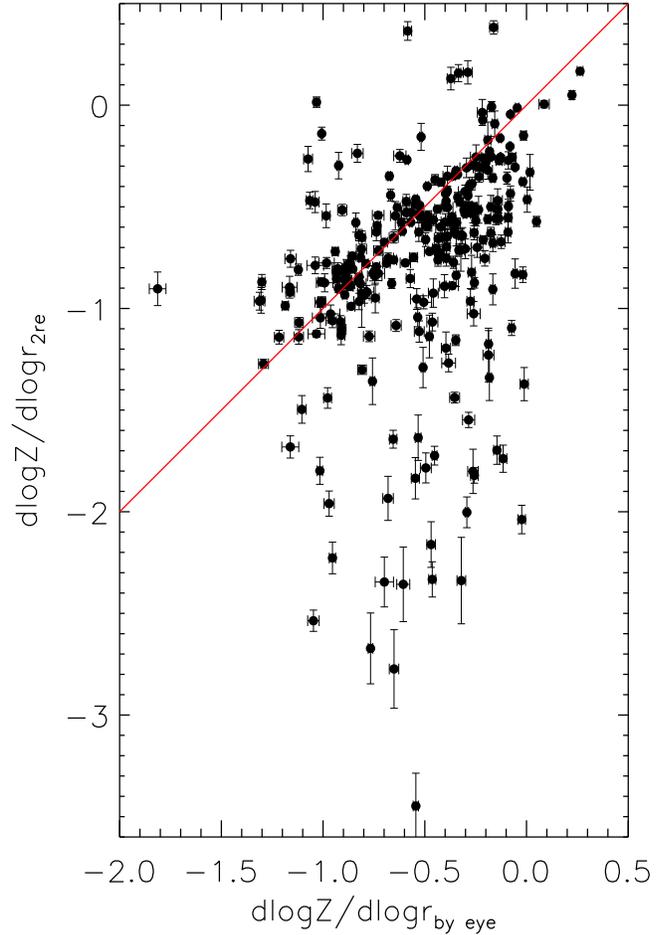}
	\caption{ {Effect of fitting range on derived metallicity gradients.
	The vertical axis shows the stellar metallicity gradient obtained using a fitting range of 2 kpc to $2R_{\rm e}$ for each galaxy.
	The horizontal axis shows stellar metallicity gradient obtained using the method described in Section \ref{sec:fit}.
	Uncertainties are estimated using the bootstrap procedure described in the main text.}}
	\label{fig:zgradcompare}
\end{figure}
 {Fig. \ref{fig:zgradcompare} compares the stellar metallicity gradients measured in galaxies from the simulation with AGN when the fitting range is fixed from $2$ kpc to $2R_{\rm e}$ (denoted ${\rm d}\log Z_*/{\rm d}\log r_{2{\rm re}}$) with the procedure described above (denoted ${\rm d}\log Z_*/{\rm d}\log r_{\rm by\,eye}$).
The solid red line denotes equality between these two quantities.
Most of the points lie on or close to this line, indicating that the precise fitting range does not affect the result for most galaxies.
There is much greater spread when a fixed range is used, particularly below the line, and large uncertainties are associated with points further from the line.
This is caused by sharply declining metallicity profiles in the outskirts of these galaxies.
}

We estimate errors on the gradient as the width of the distribution obtained from repeating the fitting to $10^5$ bootstrap realisations of each profile.
In the following sections, we show only galaxies that contain at least 1000 star particles for stellar gradients, or 1000 gas or young star particles for gas gradients.
As a result, we obtain acceptable fits for 308 and 26 galaxies for stellar and gas-phase gradients, respectively, for the simulation with AGN of $25h^{-1}$ Mpc$^3$ volume.

Fig. \ref{fig:5paper5_fit} shows how our method gives a good fit to the radial metallicity profiles of some of our simulated galaxies.
 {The galaxies shown are matched between the simulations with and without AGN, and have measured stellar and gas metallicity gradients in both.
The left-hand panels show the stellar metallicity profiles of the galaxies, and the right-hand panels show the gas-phase metallicity profiles.
Best straight-line fits (red solid lines) are also included, for the simulations with (upper panels) and without (lower panels) AGN.
The best-fitting line is plotted only in the range for which the fit was performed.
}

\subsection{Morphology Classification}
\label{sec:morph}
In observations at low redshift, ETGs are bulge-dominated (ellipticals, fast/slow-rotators, and S0), while star-forming late-type galaxies (LTGs) have disc-dominated or irregular morphologies.
For ETGs, metallicity gradients are measured with absorption lines, which are comparable to V-band luminosity-weighted stellar metallicities in our simulations.
For LTGs, metallicity gradients are measured with emission lines, which are comparable to the young stellar mass-weighted gas-phase metallicities in our simulations.
In simulations, we do not have such observational biases, and we show also the stellar metallicity gradients of LTGs in this paper.
In order to compare with observations, and also to understand the dependence of gradients on morphology, we introduce the following two quantities closely related to galaxy morphology, as well as a third that describes the galaxy environment.

\begin{figure*}
	\centering
	\subfigure{\includegraphics[width=0.49\textwidth,keepaspectratio]{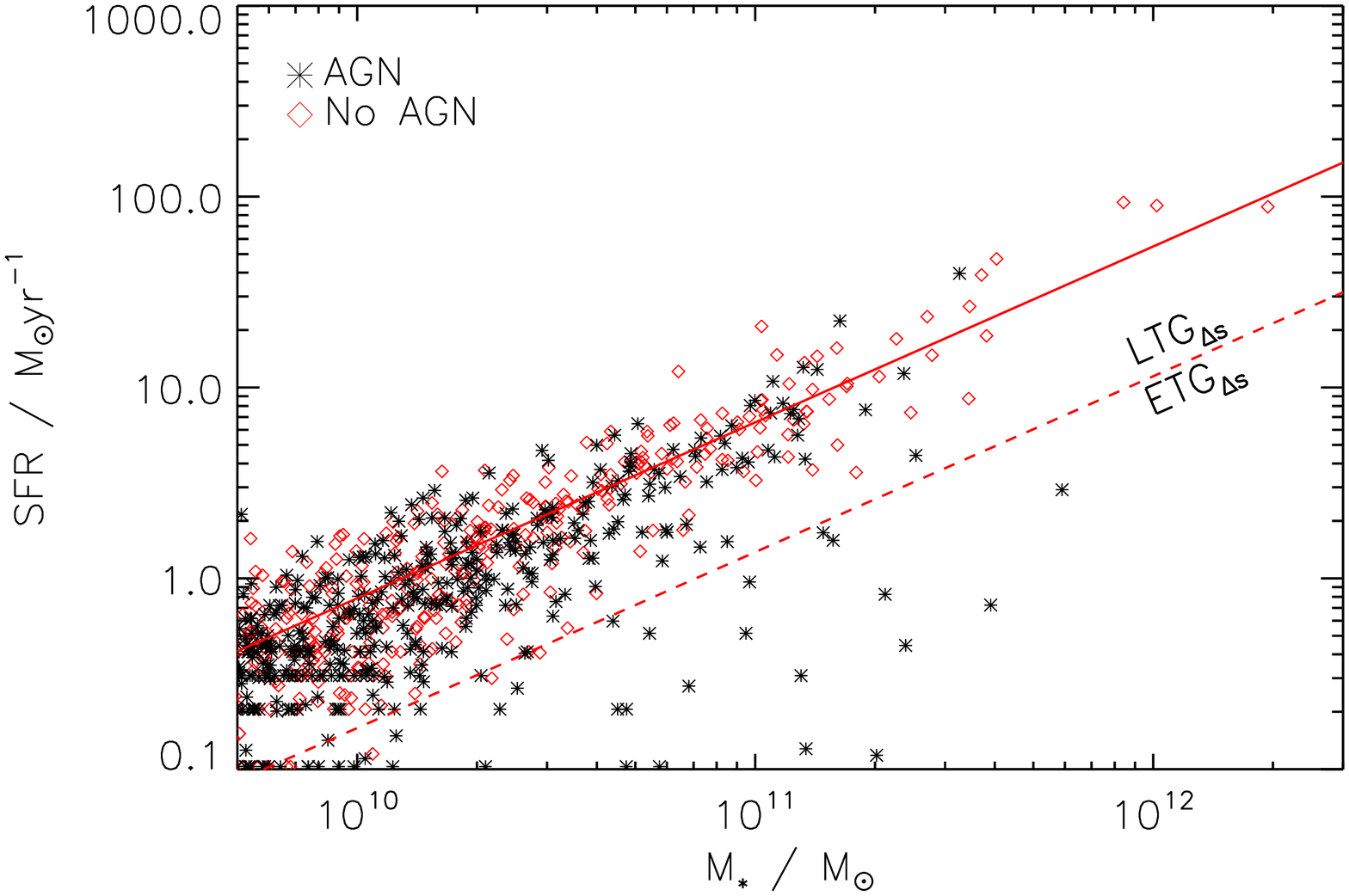}}
	\subfigure{\includegraphics[width=0.455\textwidth,keepaspectratio]{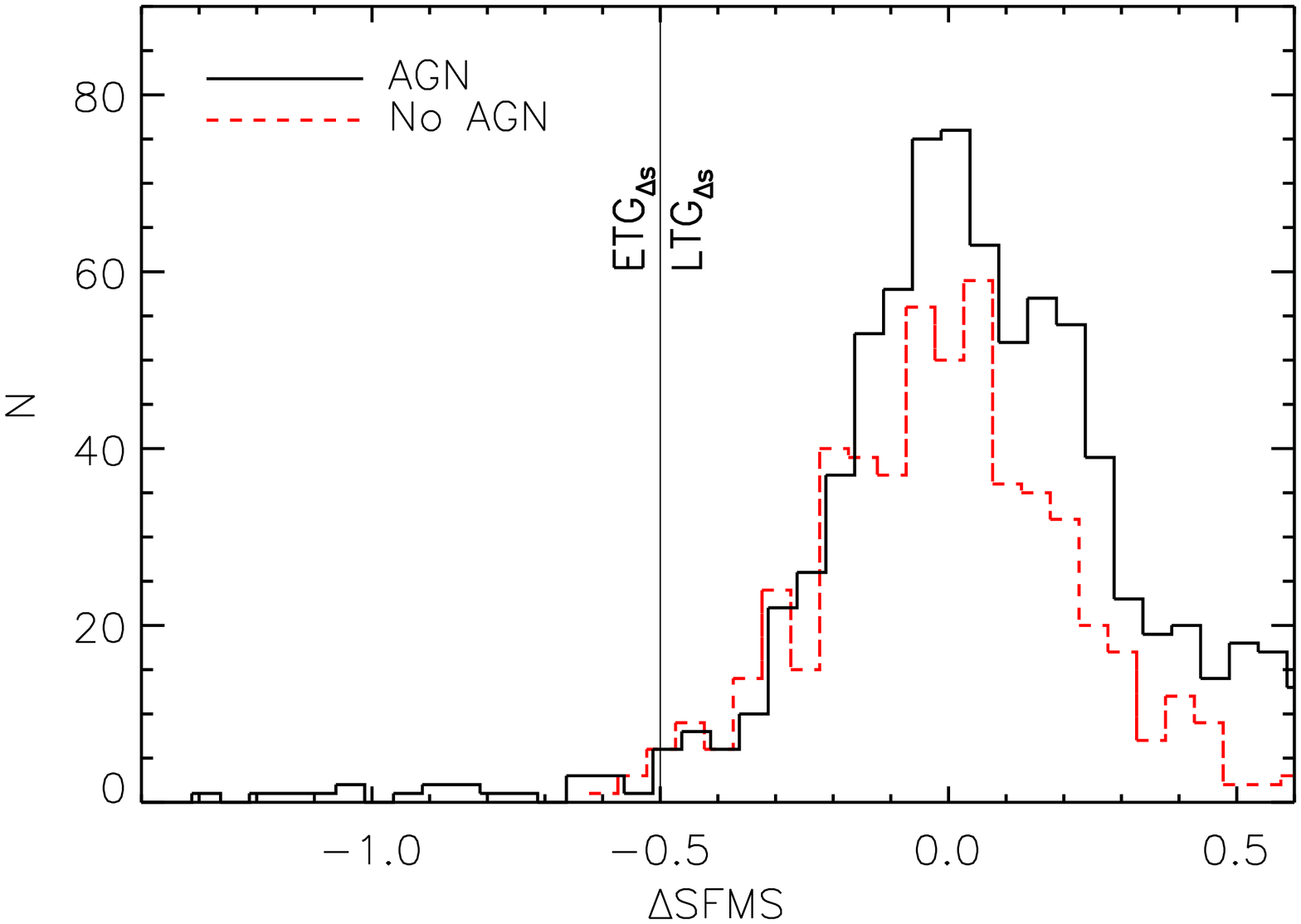}}
	\caption{{\bf Left:} SFMS from our simulations with (black stars) and without (red diamonds) AGN feedback.
	The solid red line sows the best fit to the data without AGN feedback.
	The dashed line shows this best fit line shifted down by 0.5 dex perpendicular to the line, and is the criterion of \etgs.
	{\bf Right:} Distribution of perpendicular distances ($\Delta$SFMS) from the SFMS in our two simulations (solid black line for the simulation with AGN, red dashed for the simulation without).}
	\label{fig:sfms}
\end{figure*}
Similar to colour-magnitude diagrams \citep[e.g.,][]{kauffmann03}, the so-called star-forming main sequence (SFMS) of galaxies can be used to distinguish between ETGs and LTGs in observations \citep[e.g.,][]{wuyts11,renzini15} and simulations \citep[e.g.,][]{pt16}: LTGs form the SFMS itself, while ETGs have significantly lower SFR at a given mass.
The left-hand panel of Fig. \ref{fig:sfms} shows our simulated SFMS for both simulations, as well as the best-fitting line (solid) to the data from the simulation without AGN feedback.
We define the quantity $\Delta$SFMS as the perpendicular distance of the simulated data from this line, and the distribution of these distances is shown in the right-hand panel of Fig. \ref{fig:sfms}.
In subsequent sections, we will refer to galaxies with $\Delta{\rm SFMS} >-0.5$ as \ltgs, and $\Delta{\rm SFMS}<-0.5$ as \etgs; this is shown in the left-hand panel of Fig. \ref{fig:sfms} as the dashed line.

Likely the greatest driver of morphological change is major mergers, whereby spirals can be transformed into ellipticals \citep[e.g.,][]{toomre72,ck04}.
Therefore, we construct merger trees for each galaxy (see Appendix \ref{ap:tree} for the details) in the simulation at $z=0$ and trace their evolution back to their most recent major merger (defined as having a mass ratio greater than $1/4$).
This time difference, $t_{\rm merge}$ can then give an indication of morphology, with smaller $t_{\rm merge}$ more likely to be bulge-dominated.
In subsequent sections, we will refer to galaxies with $t_{\rm merge}>1$ Gyr as \ltgt, and $t_{\rm merge}<1$ Gyr as \etgt.
Note that it is possible to have bulge-dominated morphology at $t_{\rm merge} \sim 1-10$ Gyr if there is no significant gas accretion after the last major merger, and thus \etgt may be a subset of elliptical galaxies.

\begin{figure}
	\centering
	\includegraphics[width=0.48\textwidth,keepaspectratio]{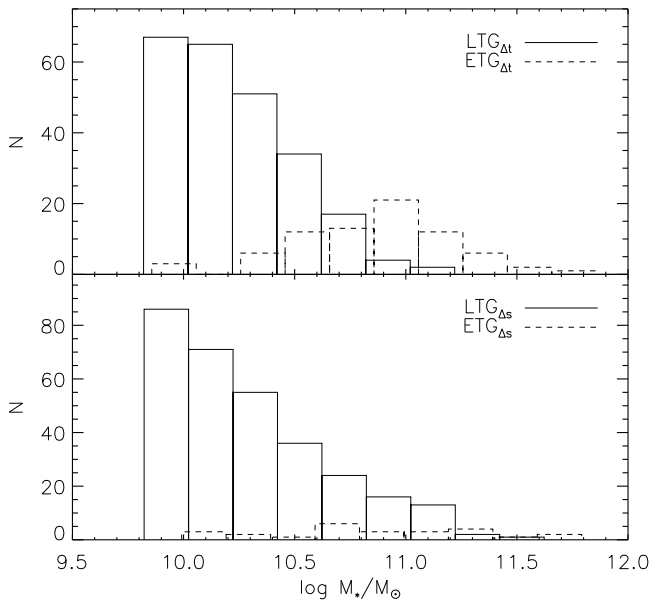}
	\caption{{\bf Top panel}: mass distributions for galaxies defined as either \ltgt (solid lines) or \etgt (dashed lines).
	{\bf Bottom panel}: mass distributions for galaxies defined as either \ltgs (solid lines) or \etgs (dashed lines).
	Only galaxies from the simulation with AGN feedback are shown.}
	\label{fig:morphproxies}
\end{figure}

 {It is important to note that we have not classified our simulated galaxies by morphology itself in this paper, and more detailed analysis of the morphology, including kinematical decomposition, will be made in future work.
In Fig. \ref{fig:morphproxies} we show the distributions of our morphology indicators, \etgs, \ltgs, \etgt, and \ltgt, with stellar mass.
The top panel shows that  \ltgt dominate at low mass,  and \etgt, which have experienced a major merger within 1 Gyr, are more numerous above $M_*\sim 5\times10^{10}\msun$; the \etgt fraction is $\sim 45$ per cent and $\sim85$ per cent at $M_*\sim 5\times10^{10}\msun$ and $\sim 10^{11}\msun$, respectively.
Since mergers are strong drivers of morphological change, this implies that our high-mass galaxies are mostly elliptical, which is in good agreement with observational results for ellipticals and S0 galaxies \citep[e.g.,][]{bernardi10,tempel11}.
There is not a perfect correlation between galaxy morphology and $t_{\rm merge}$, however, and the details of each individual merger are important.
In the lower panel of Fig. \ref{fig:morphproxies}, galaxies are separated by $\Delta$SFMS.
Galaxies that lie close to the SFMS (\ltgs) are most numerous at lower masses, and constitute a declining fraction of the galaxy population with increasing mass, but the fraction of \etgs is not perfectly consistent with observations \citep[e.g.,][$\sim 50$ per cent and $\sim 75$ per cent at $\log M_* = 10.5$ and 11]{renzini15}.
This means that our AGN feedback is probably not strong enough (see \citetalias{pt15a} for more discussion).
Our predicted distribution of gradients could be affected by these effects.
Our galaxies that are classified as both \ltgs and \etgt should be ellipticals, and their metallicity gradients may be too steep because of the lack of quenching of star formation. 
Therefore, we think our sample of \etgs is better for the comparison with observations of SAURON and ATLAS$^{\rm 3D}$.
For low-mass galaxies, the lack of resolution could result in underestimating gradients in general.
However, our low-mass \ltgs and \ltgt galaxies, it could also overestimate the gradients in the case that the gradients would be measured along small star-forming discs, which are not resolved in our simulations.
}

\begin{figure}
	\centering
	\includegraphics[width=0.48\textwidth,keepaspectratio]{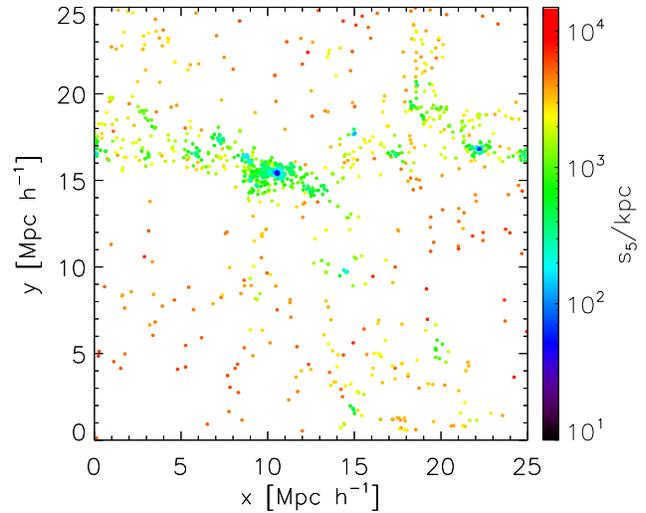}
	\caption{The distribution of galaxies in the simulation with AGN feedback, coloured by $5^{\rm th}$ nearest neighbour distance, $s_5$.}
	\label{fig:s5}
\end{figure}
\begin{figure}
	\centering
	\includegraphics[width=0.48\textwidth,keepaspectratio]{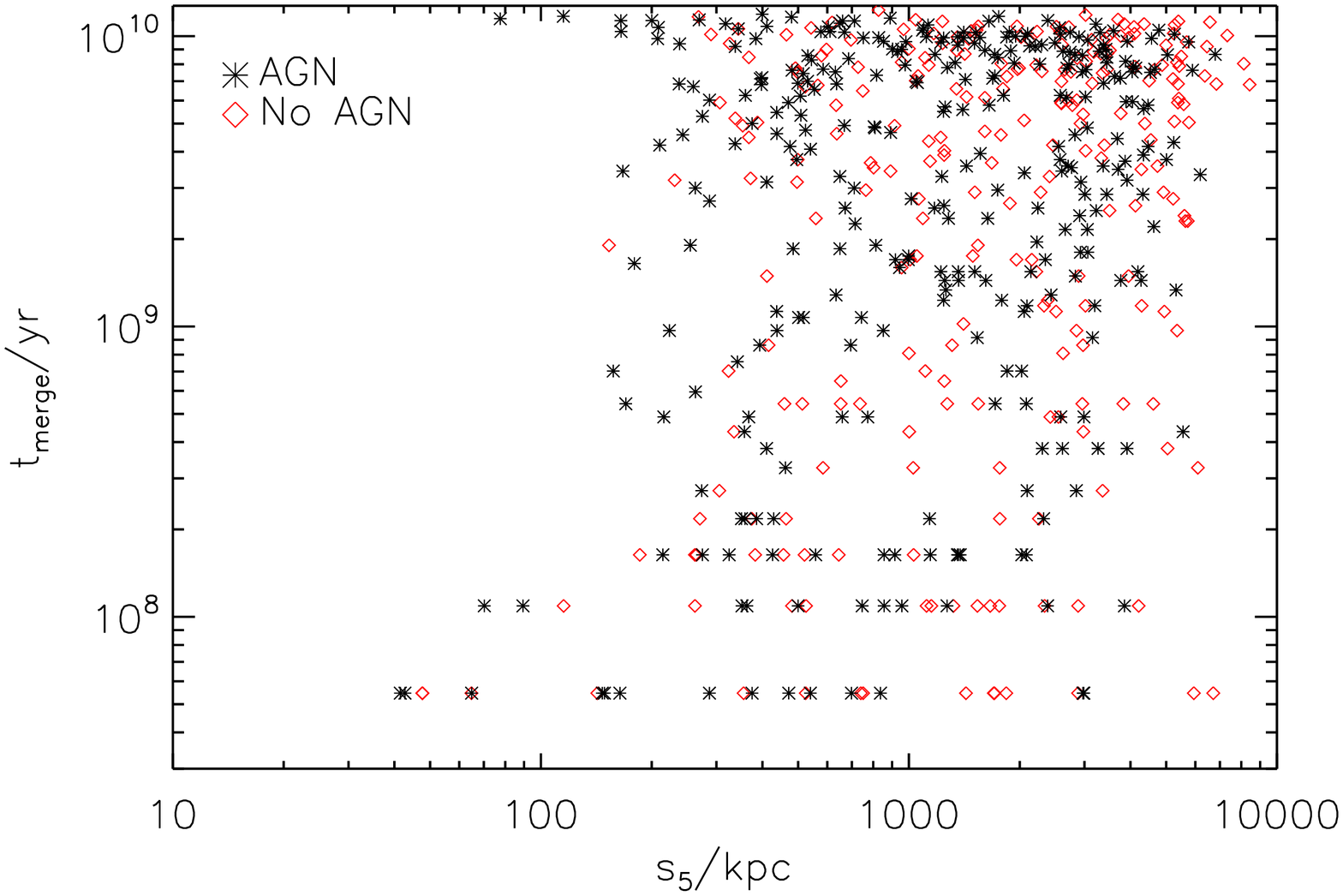}
	\caption{Time since a galaxy last experienced a major merger, $t_{\rm merge}$ as a function of $5^{\rm th}$-nearest neighbour distance, $s_5$.
	In the densest environments (low $s_5$), almost all galaxies have experienced a major merger in the recent past ($\ll1$ Gyr).
	Black stars and red diamonds represent galaxies from the simulations with and without AGN feedback, respectively.}
	\label{fig:s5tmerge}
\end{figure}
We also introduce a measure of galaxy environmental density.
It is well known that galaxy morphology depends on environmental density, with high density environments like clusters having a higher fraction of elliptical galaxies, while field galaxies are mostly late-types \citep{dressler80,cappellari11,houghton15}.
Note that the morphology--density relation is not a one-to-one correspondence; for example, E+A galaxies can be located outside of clusters \citep[e.g.,][]{zabludoff96}.
We use the 3-D $5^{\rm th}$-nearest neighbour distance, $s_5$ as our measure of environmental density, with low $s_5$ corresponding to high density (see Appendix \ref{ap:s5} for a discussion on the choice of $n=5$ for $s_n$).
Fig. \ref{fig:s5} shows the distribution of galaxies from one of our simulations coloured by $s_5$.
We also show, in Fig. \ref{fig:s5tmerge} the ``morphology--density" relation from our simulations, with $t_{\rm merge}$ as a proxy for morphology.
In high density environments, most galaxies are \etgt and have undergone recent mergers, suggesting an elliptical morphology, whereas a greater fraction of galaxies with large $s_5$ have not experienced a merger in many Gyrs.
Note that the correlation between $\Delta$SFMS and $s_5$ is weaker than Fig. \ref{fig:s5tmerge}, although galaxies tend to be \etgs with low negative $\Delta$SFMS in high density environments.

 {
The four galaxies shown in Fig. \ref{fig:5paper5_fit} are \ltgs in both simulations.
Galaxy ga0025 is \ltgt in the simulation without AGN ($t_{\rm merge}=1.5\times10^9$ yr), otherwise the galaxies are \etgt.
These galaxies exist in a range of environments, having $150\ltsim s_5\ltsim2,250$ kpc.
}


\section{Stellar Metallicity Gradients}\label{sec:5paper5_starz}
\subsection{Gradients vs Mass}\label{sec:5paper5_starz0}

\begin{figure}
	\centering
	\includegraphics[width=0.48\textwidth,keepaspectratio]{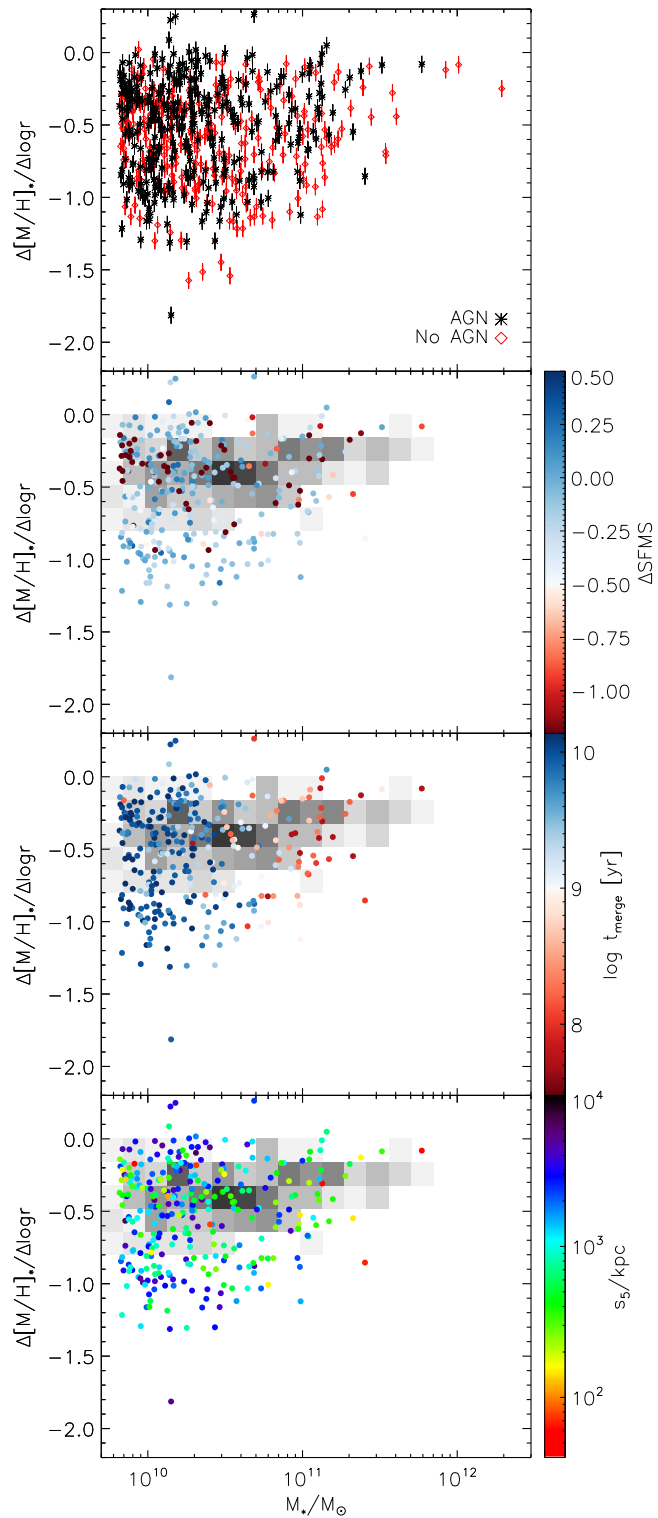}
	\caption{{\bf Top panel:} Stellar metallicity gradient as a function of galaxy mass at $z=0$ for galaxies in the simulation with (black) and without (diamonds) AGN feedback.
	{\bf Second, third, and fourth panels:} Data from the simulation with AGN feedback, coloured by $\Delta$SFMS, $t_{\rm merge}$, and $s_5$, respectively.
	The grey shaded area shows the region occupied by ETGs from the ATLAS$^{\rm 3D}$ survey (Kuntschner et al., in prep).}
	\label{fig:starz}
\end{figure}
In this section, we present the present-day stellar metallicity gradients of our simulated galaxies.
The top panel of Fig. \ref{fig:starz} shows the gradients as a function of galaxy stellar mass for our simulations with (black stars) and without (red diamonds) AGN feedback.
Bootstrap error bars are also shown.
Most of our simulated galaxies have negative metallicity gradients, meaning that their centres show greater chemical enrichment than the outskirts.
More massive galaxies tend to have flat (close to 0) slopes, while at lower masses there is a much greater range.
We investigate the origin of this variation further in the lower panels, for the case with AGN feedback.

\begin{figure}
	\centering
	\includegraphics[width=0.48\textwidth,keepaspectratio]{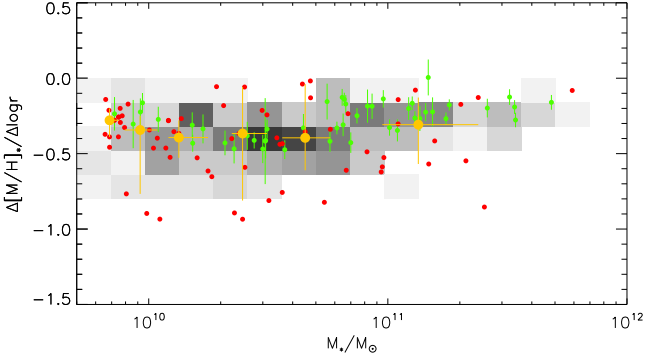}
	\caption{Stellar metallicity gradient as a function of mass for those simulated galaxies classified as \etgs in the simulation with AGN feedback (red points).
	 {Large orange points show the median simulated relation.}
	Observational data from the SAURON \citep[green points with error bars,][]{kuntschner10} and ATLAS$^{\rm 3D}$ (grey shading, Kuntschner et al., in prep.) surveys are also shown.}
	\label{fig:starz_sfms_etg}
\end{figure}

In the second panel of Fig. \ref{fig:starz}, we colour the data by $\Delta$SFMS.
Galaxies that lie close to the SFMS (classified as \ltgs) have steep metallicity gradients, and dominate the distribution of gradients at $\Delta$[M/H]$_*/\Delta\log r \ltsim -0.8$.
This is because star formation concentrated towards the centre of galaxies steepens the gradient with time.
On the other hand, \etgs have significantly flatter gradients at all masses, which follow the observational data perfectly.
The observationally-derived gradients from the ATLAS$^{\rm 3D}$ survey are shown by the grey shaded region, with a greater density of galaxies indicated by a darker colour.
To make a clearer comparison, we show in Fig. \ref{fig:starz_sfms_etg} just simulated \etgs (red points), along with observational data from the SAURON survey \citep[green points with error bars,][]{kuntschner10} and ATLAS$^{\rm 3D}$ survey\footnote{ {SAURON data extend to approximately $1 R_{\rm e}$ \citep{kuntschner10}. This is very similar the typical range used for fitting our simulated data, and so a meaningful comparison can be made.}}.
The turnover in gradient identified by \citet{spolaor09,spolaor10} at $\sim 3\times10^{10}\msun$ is clearly seen in both observational datasets, and also exists in our simulated \etgs.
 {The large orange points show the median simulated relation, which is in excellent agreement with the observational data.
We further note that a similar turnover was seen in the metallicity gradients of simulated spiral galaxies in \citet{tissera16}, where a different prescription for SN feedback was used, and AGN feedback was not included.}

In the hierarchical model of galaxy growth, more massive galaxies are more likely to have experienced a major merger in their past; the third panel of Fig. \ref{fig:starz} displays the same data, coloured by the last major merger epoch, $t_{\rm merge}$, and it is clear that this prediction from the hierarchical model is borne out in our simulations.
Most galaxies with masses $\gtsim 5\times 10^{10}\msun$ are \etgt (see Fig. \ref{fig:morphproxies}), having had a major merger within the previous 1 Gyr.
 {This explains the overall trend of flatter gradients with galaxy mass since major mergers act to redistribute the stars of the merging galaxies, somewhat homogenizing the distribution of metals and flattening the gradient.
 {As noted in Section \ref{sec:morph}, some of the steep gradients of galaxies that are both \ltgs and \etgt may be due to the lack of suppression of gradient regeneration.}
There is still scatter at a given mass since the details of the merger, in particular the gas mass of the galaxies, is also important.
Following wet mergers (both galaxies are gas-rich) strong central star formation may occur and re-steepen the gradient \citep{ck04}.
On the other hand, dry mergers provide no new fuel for star formation.
This is true regardless of whether or not AGN feedback is included, and the prediction of \citet[][the evolutionary path iii in Section \ref{sec:intro}]{ck04}, which did not include AGN feedback, stands.}

 {For $M_*\gtsim3\times10^{10}\msun$, galaxies that are both \ltgs and \ltgt are likely to be disc-dominated, and show much smaller variation in mass, which is consistent with the finding by \citet{sanchez14}.
At lower mass, however, there are \ltgs that have not undergone any major merger but do have shallow metallicity gradients.
This could be due to supernova feedback; in this low-mass end, supernova feedback can drive galactic winds to quench star formation.
If chemical enrichment takes place at the front of such winds, the metallicity gradients can be flat or even positive \citep{mori97}.}

The final panel of Fig. \ref{fig:starz} shows the environmental dependence, i.e., the mass -- metallicity gradient relation coloured by $s_5$, the $5^{\rm th}$-nearest neighbour distance.
Through the morphology--density relation \citep{dressler80}, $s_5$ may be the best proxy for galaxy morphology available for our simulated galaxies.
 {\etgs tend to be found in denser environments (median $s_5=366$ kpc) compared to \ltgs (median $s_5=1,507$ kpc).}
 {Reflecting this morphological dependence, there is a weak trend of metallicity gradient with $s_5$; the galaxies with $s_5<200$ kpc tend to have flatter gradients at a given mass, and the two exceptions with $s_5<200$ kpc and steep gradients ($\Delta$[M/H]$_*/\Delta\log r \sim -1$) are in the green valley in our definition ($\Delta{\rm SFMS}\sim-0.5$).}
Indeed, most simulated galaxies with $s_5\ltsim150$ kpc are consistent with the ATLAS$^{\rm 3D}$ data, whose samples are ETGs including Virgo cluster members,  {and all are \etgs.}
 {Among \etgs, however, we find no significant environmental dependence because the gradients depend more strongly on the details of each individual merger (e.g., gas fractions), rather than environment itself.
Galaxies in high-density environments can also be fed by cold streams from the cosmic web, so the dependence on environment could potentially be very complicated.}

\begin{figure}
	\centering
	\includegraphics[width=0.48\textwidth,keepaspectratio]{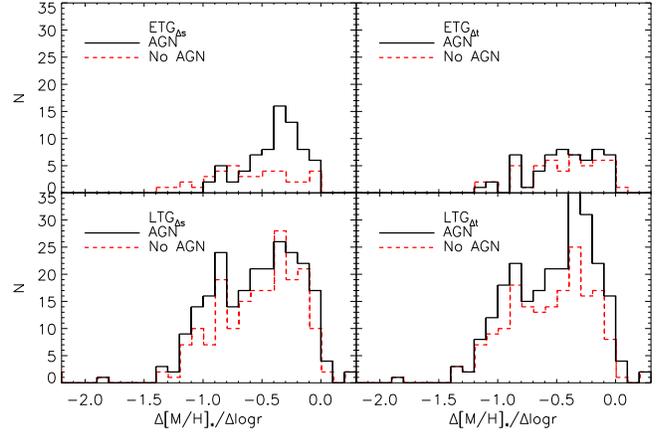}
	\caption{Present-day stellar metallicity gradient distributions for \etgs (top left), \etgt (top right), \ltgs (bottom left), and \ltgt (bottom right).
	Solid black histograms denote the simulation with AGN, and dashed red without.}
	\label{fig:starzhist}
\end{figure}

In the first panel of Fig. \ref{fig:starz} there may be a small difference between the simulations with and without AGN feedback.
To provide a more quantitative comparison, we divide the galaxies into early- and late-type, using our proxies for morphology, and compare the resulting distributions, which are shown in Fig. \ref{fig:starzhist}.
In the top left panel, comparing ETGs defined by their proximity to the SFMS, the distribution when AGN is included seems shifted to flatter gradients than when AGN feedback is not included.
 {This is because AGN feedback quenches star formation in massive galaxies earlier than would otherwise occur, after which they are unable to steepen their gradient via central star formation, and are therefore more susceptible to the effects of mergers.
Since quenching occurs around $z\sim2$, such mergers can happen any time in the last $\sim10$ Gyr, which is why the populations defined as \etgt do not show any distinction in their present-day distributions.
 {As noted in Section \ref{sec:morph}, the difference in the distributions for \etgs and \etgt is caused by the lack of suppression of gradient regeneration for the galaxies classified as both \ltgs and \etgt.
The difference between \ltgs and \ltgt may be caused by the unresolved discs with flat gradients.}
}

The median of metallicity gradients for \etgs and \etgt is $\Delta$[M/H]$_*/\Delta\log r = -0.37$ and $-0.42$, respectively.
These are slightly steeper than in the simulations of \citet{ck04}.
This is because gas accretion continues and the star formation timescale is longer in our simulations than in \citet{ck04}.
Even so, we do not find the gradients in very massive galaxies to be as steep as in the observations of brightest cluster galaxies \citep{brough11,loubser11}.
The origin of such steep gradients could be gas accretion.

\begin{figure}
	\centering
	\includegraphics[width=0.48\textwidth,keepaspectratio]{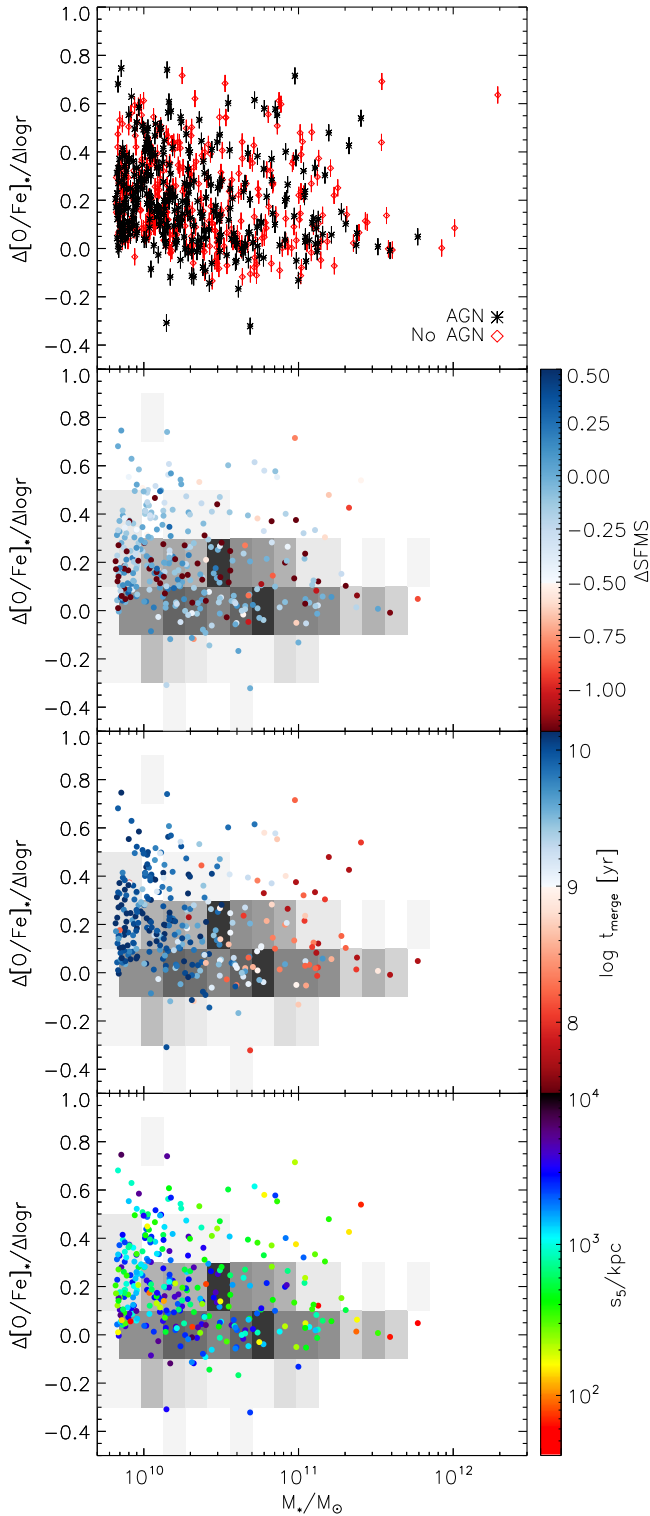}
	\caption{{\bf Top panel:} Stellar [O/Fe] gradient as a function of galaxy mass at $z=0$ for galaxies in the simulation with (black) and without (diamonds) AGN feedback.
	{\bf Second, third, and fourth panels:} Data from the simulation with AGN feedback, coloured by $\Delta$SFMS, $t_{\rm merge}$, and $s_5$, respectively.
	The grey shaded area shows the region occupied by ETGs from the ATLAS$^{\rm 3D}$ survey (Kuntschner et al., in prep).}
	\label{fig:starofe}
\end{figure}
\begin{figure}
	\centering
	\includegraphics[width=0.48\textwidth,keepaspectratio]{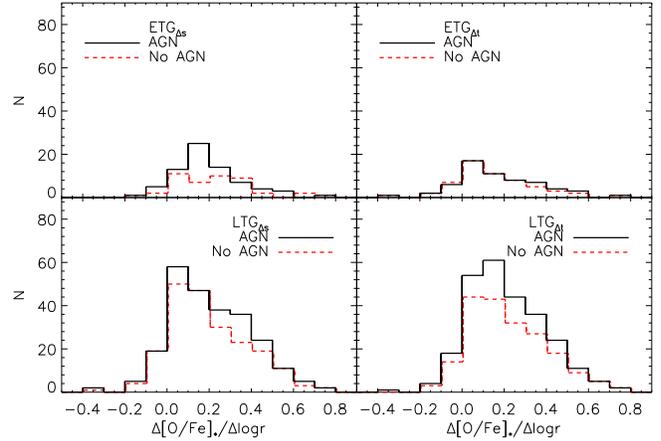}
	\caption{Present-day stellar [O/Fe] gradient distributions for \etgs (top left), \etgt (top right), \ltgs (bottom left), and \ltgt (bottom right).
	Solid black histograms denote the simulation with AGN, and dashed red without.}
	\label{fig:starofe_hist}
\end{figure}

We also show, in Fig \ref{fig:starofe}, the gradient in [O/Fe]$_*$ for the galaxy population of our two simulations at $z=0$.
On average, simulated galaxies have a small positive gradient ($\sim0.1$ - 0.2), which means [O/Fe]$_*$ is slightly lower in the centre than the outskirts.
This is because star formation takes place over longer timescales towards the centre of the galaxy due to the deeper gravitational potential well and there is greater contribution from Type Ia supernovae.
Note that Fe is produced more in SNe Ia relative to O, with a timescale that depends on metallicity spanning a range $0.2-20$ Gyr at solar metallicity in our model \citep[see][for more details]{ck09}.
In massive galaxies, the gradient is generally shallower, which is due to the quenching of star formation in the centre before the majority of Type Ia contribute.
This is confirmed in the first panel, where the simulation without AGN feedback has larger [O/Fe]$_*$ gradients in massive galaxies.
 {These galaxies have ongoing central star formation at late times, from gas that has been polluted by previous generations of SNe Ia.
However, with AGN, [O/Fe]$_*$ gradients of massive galaxies become almost 0.}

In the second panel, as was the case for metallicity gradients, galaxies with steep [O/Fe]$_*$ gradients are often star-forming (\ltgs), while \etgs are in good agreement with ETGs from the ATLAS$^{\rm 3D}$ survey (grey shading).
 {
In the third panel, there is no difference in the distributions of gradients defined by the time since last major merger, $t_{\rm merge}$.
Although major mergers act to redistribute metals and flatten the gradient, the details of the merger are important, and merger-induced central star formation can steepen gradients.
This leads to the large scatter seen in the \etgt population (see also Section \ref{sec:morph}).
In the last panel, the data are coloured by $5^{\rm th}$-nearest neighbour distance, $s_5$.}
 {As for the metallicity gradients, there is only a weak trend reflecting the morphology--density relation, whereby galaxies with $s_5<200$ kpc tend to show no [O/Fe]$_*$ gradients.
However, there is no clear trend for the population as a whole.}

We again bin galaxies by morphological type, as defined by $\Delta$SFMS and $t_{\rm merge}$, and the resulting distributions are shown in Fig. \ref{fig:starofe_hist}.
As for the metallicity gradients, only the top left panel shows a significant difference in the distributions, where [O/Fe]$_*$ gradients of \etgs are flatter with AGN feedback.

\subsection{Time Evolution}
\label{sec:5paper5_res_gradt}
\begin{figure*}
\centering
\includegraphics[width=\textwidth,keepaspectratio]{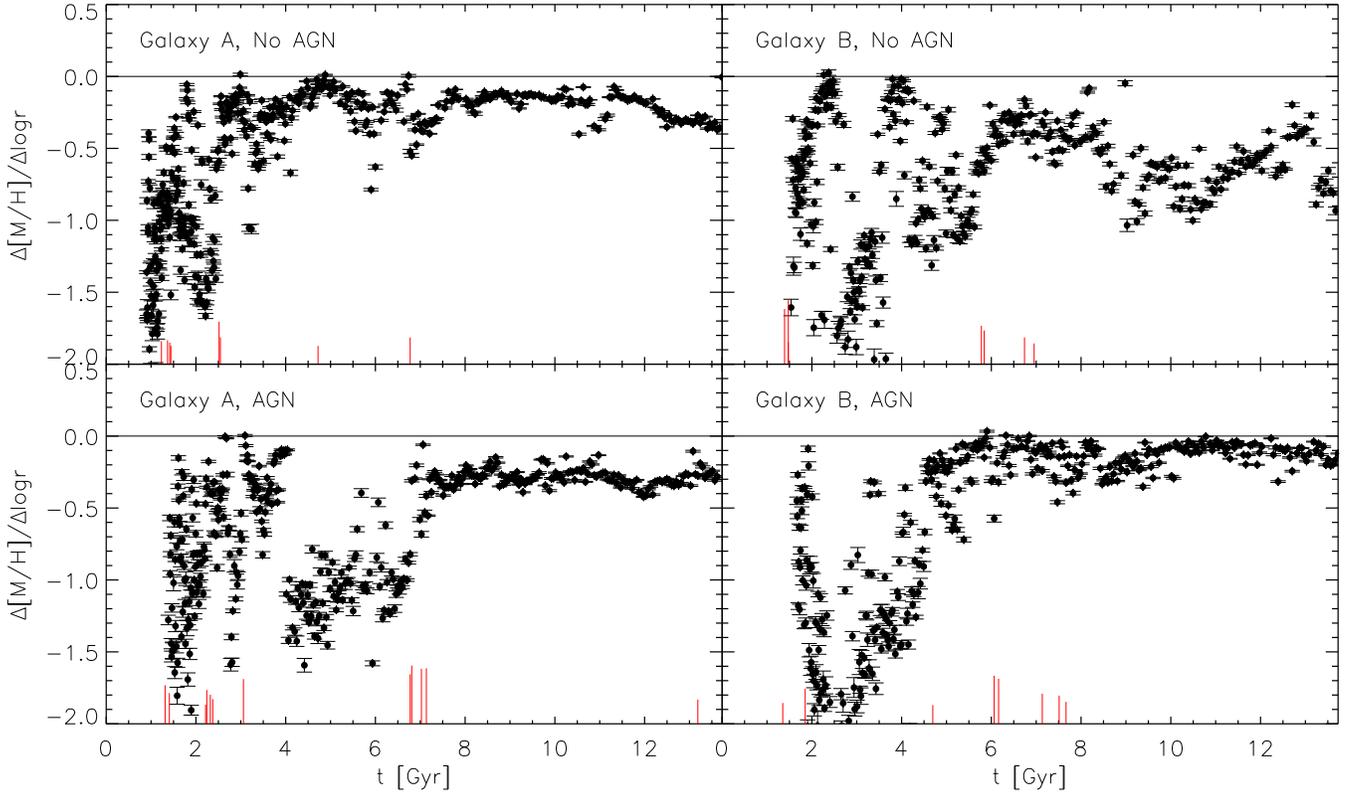}
\caption{Stellar metallicity gradients in the galaxies A (left panels) and B (right panels) across cosmic time both with (lower panels) and without (top panels) AGN.
Vertical lines show the approximate times that these galaxies undergo a merger with mass ratio 1:4 or greater.
The height of the lines is proportional to the mass ratio, with a 1:1 merger reaching to the first major tick mark on the ordinate axis.}
\label{fig:5paper5_gradtime}
\end{figure*}

We highlight some of the results of the previous section by explicitly following the evolution of the metallicity gradient of two massive galaxies, labelled A and B, matched between the two simulations.
A is the most massive galaxy at $z=0$ and sits at the centre of a cluster; B is a field galaxy with a comparatively quiet merging history (see \citetalias{pt15a} for more details, and Appendix \ref{ap:tree} for the merger trees).
Fig. \ref{fig:5paper5_gradtime} shows the evolution of the metallicity gradient in these galaxies as a function of cosmic time.
Also shown, by the vertical lines, are the approximate times of all mergers onto these galaxies with a stellar mass ratio of at least 1:4.
The height of each line is proportional to the mass ratio of the merger, with a 1:1 merger represented by a line reaching to the first major tick on the ordinate axis.

At early times ($t\ltsim6-7$ Gyr; $z\gtsim1$), the metallicity gradients of both galaxies in both simulations are strongly affected by successive gas-rich mergers and high rates of star formation.
This assembly of gas-rich subgalaxies is a similar situation to the monolithic collapse (the evolutionary path i in Section \ref{sec:intro}), which results in gradients as steep as $\Delta$[M/H]$_*/\Delta\log r \sim -1.5$.
However, gradients in this period show greater spread than at later times, which is caused by mergers, different from the monolithic collapse scenario \citep{pipino10}.
As the galaxies form and grow in size, the gradient initially becomes shallower as the outskirts of the galaxies become more enriched by star formation and/or stellar accretion (evolutionary path ii in Section \ref{sec:intro}).
Subsequently, ongoing central star formation can cause the metallicity gradients to steepen, especially in the simulation without AGN feedback since the star formation rate stays high for the lifetime of the galaxies {; this is especially clear in the evolution of the gradient  {(from -0.2 to -0.7)} in galaxy B at late times ($t\gtsim6$ Gyr).}

Major mergers tend to cause gradients to become shallower, as metal-rich stars from the centre of the galaxy are suddenly redistributed further out.
This is shown most clearly immediately following the triple galaxy merger, with similar-mass components, that galaxy A experiences in both simulations at $t\sim6.75$ Gyr {; the gradient evolves from -1 to -0.3 for the case with AGN.}

This event is highly disruptive to the structure of all three components, and the gradient becomes significantly shallower as a result.
In the simulation with AGN, this merger also triggers the growth of this galaxy's BH, and star formation is suppressed by AGN feedback by an order of magnitude or more for the remainder of its life (see Fig. 4 of \citetalias{pt15a}).
Therefore, the gradient is not able to steepen again, unlike in the simulation without AGN  {($t\gtsim 12$ Gyr for both galaxies)}.

We also note that when AGN feedback is included, the gradients at late times ($t\gtsim6$ Gyr; $z\ltsim1$) show less scatter between snapshots than for the corresponding galaxies without AGN.
By $z\sim1$, the black hole at the centre of these galaxies has grown sufficiently massive to effectively suppress star formation.
With stochastic star formation away from the galaxy centre, the metallicity gradient therefore remains almost unchanged.

\section{Gas Metallicity Gradients}
\label{sec:5paper5_massmet}
\begin{figure}
	\centering
	\includegraphics[width=0.48\textwidth,keepaspectratio]{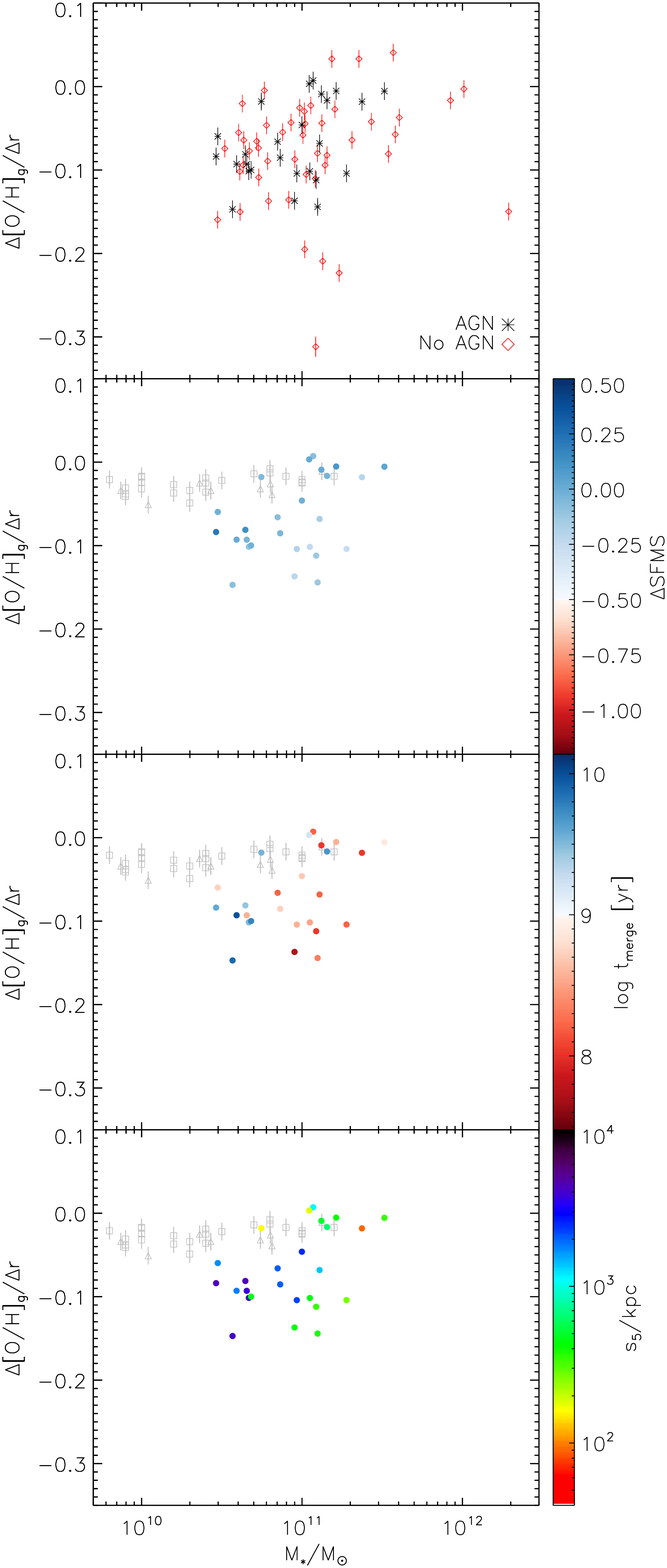}
	\caption{{\bf Top panel:} Gas-phase oxygen abundance gradient as a function of galaxy mass at $z=0$ for galaxies in the simulation with (black) and without (diamonds) AGN feedback.
	{\bf Second, third, and fourth panels:} Data from the simulation with AGN feedback, coloured by $\Delta$SFMS, $t_{\rm merge}$, and $s_5$, respectively.
	Also shown are the observational data of \citet[][grey squares]{ho15} and \citet[][grey triangles; galaxy masses were obtained from \citealt{mcgaugh05,mcgaugh15}]{kudritzki15}.}
	\label{fig:gasz}
\end{figure}

Understanding the evolution of gas in galaxies is of great importance since it is the fuel for star formation and black hole growth, and the medium in which metals are transported both within and outside a galaxy.
Indeed, the origin of gas in ETGs is the subject of ongoing uncertainty \citep[e.g.,][]{young11,lagos14}.
Compared to stellar metallicity gradients, the advantage is that gas-phase metallicity gradients can be observed out to higher redshift than can stellar metallicities \citep{cresci10,jones10,yuan11}.
In order to predict the gas-phase metallicity, it is necessary to include cosmological inflows and outflows from AGN feedback, as in our simulations \citep{pt15b}.

The top panel of Fig. \ref{fig:gasz} shows the distribution of gas-phase oxygen abundance gradients with galaxy stellar mass at $z=0$ for our simulations with and without AGN feedback (black stars and red diamonds, respectively).
The simulations with and without AGN feedback produce fairly consistent results for intermediate mass galaxies, having most gradients in the range $-0.2< \Delta{\rm [O/H]}_{\rm g}/\Delta r<0$; the simulation without AGN has galaxies to higher stellar masses because these galaxies have had their star formation quenched in the simulation with AGN.

In the second panel of Fig. \ref{fig:gasz} we show the data from the simulation with AGN feedback, coloured by $\Delta$SFMS.
All of the simulated galaxies included here are classified as \ltgs due to weighting by young stellar mass (Section \ref{sec:fit}).
Also shown are the observational data of \citet[][grey squares]{ho15}, who measured metallicity gradients along deprojected radius in local, star-forming field galaxies, and \citet[][grey triangles; galaxy masses were obtained from \citealt{mcgaugh05,mcgaugh15}]{kudritzki15}.
 {At this mass range,} the observational data follow a much tighter trend of abundance gradient with galaxy mass\footnote{At lower galaxy mass, $10^8\ltsim M_*/\msun\ltsim10^9$, the observed [O/H] gradients become steeper $\sim-0.3$ dex kpc$^{-1}$.} than simulated galaxies, and are also flatter in other works; $\Delta$[M/H]$_{\rm g}/\Delta r \sim -0.05$ dex/kpc \citep{zaritsky94} and $-0.041$ dex/kpc \citep{rupke10}.
The Milky Way Galaxy also has $\sim -0.05$ dex/kpc \citep{maciel06}.
 {Our simulated galaxies with $\Delta{\rm SFMS}>0$ overlap these observations, but others have somewhat steeper gradients, which may not be included in the sample of the observations.}
  {As noted in Section \ref{sec:morph}, for low-mass galaxies, since we cannot resolve star formation on discs, the metallicity gradients may be overestimated.}

As for stellar metallicity gradients, there is no significant relationship between the gas-phase metallicity gradients and mass.
The origin of this scatter is unclear; there is no correlation between gradient and the merging history ($t_{\rm merge}$), and only a weak correlation with environment ($s_5$) (third and fourth panels of Fig. \ref{fig:gasz}, respectively).
The lack of environmental effect was also reported in the set of simulations  {of disc galaxies} by \citet{few12}.

\begin{figure}
	\centering
	\includegraphics[width=0.48\textwidth,keepaspectratio]{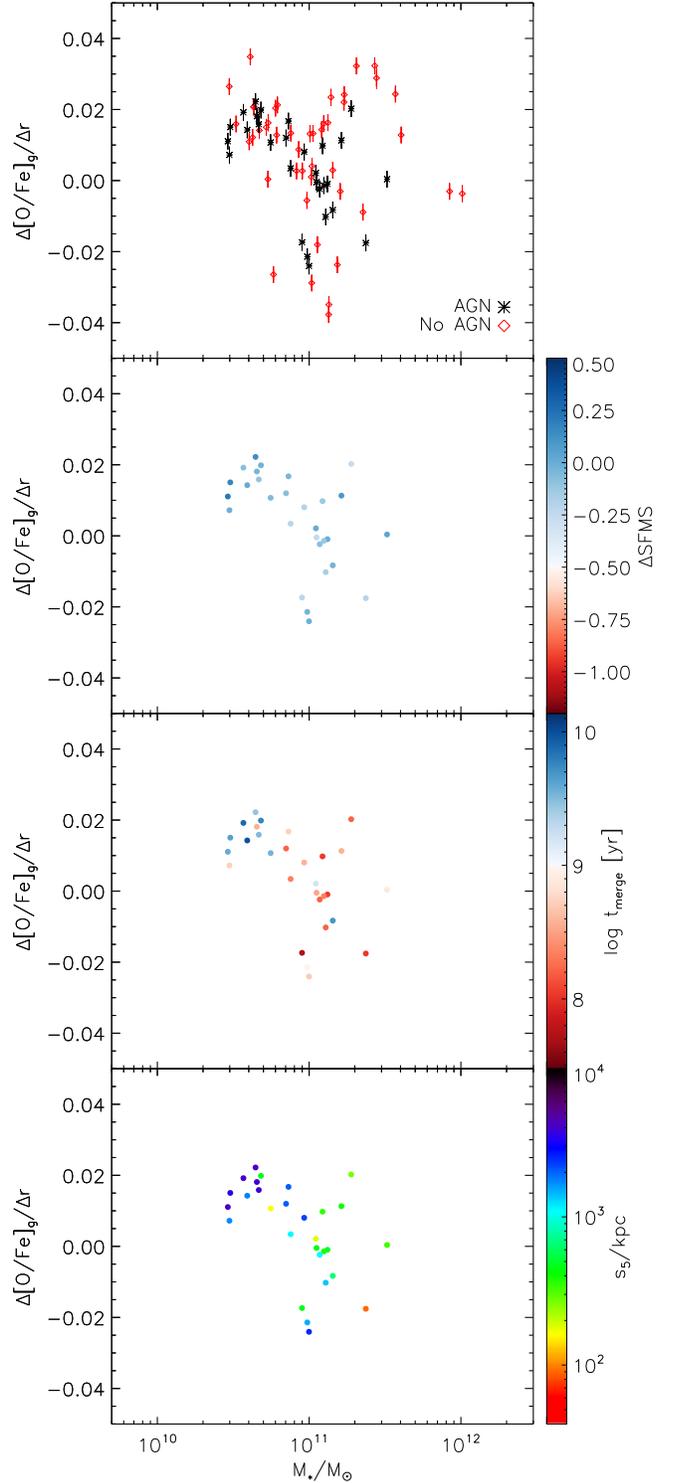}
	\caption{{\bf Top panel:} Gas-phase [O/Fe] gradient as a function of galaxy mass at $z=0$ for galaxies in the simulation with (black) and without (diamonds) AGN feedback.
	{\bf Second, third, and fourth panels:} Data from the simulation with AGN feedback, coloured by $\Delta$SFMS, $t_{\rm merge}$, and $s_5$, respectively.}
	\label{fig:gasofe}
\end{figure}

The top panel of Fig. \ref{fig:gasofe} shows the distribution of gas-phase [O/Fe] gradients as a function of stellar mass at $z=0$ for our simulated galaxies in both simulations.
In both cases the gradients are very close to flat, having values $-0.02\ltsim\Delta{\rm [O/Fe]}_{\rm g}/\Delta r\ltsim0.02$ dex kpc$^{-1}$.
There is a weak correlation in the simulation with AGN feedback whereby less massive galaxies have larger values of gradients  {because of the quenching of iron enrichment at the centre}.
The lower three panels of Fig. \ref{fig:gasofe} show the data from the simulation with AGN feedback coloured by $\Delta$SFMS, $t_{\rm merge}$, and $s_5$, respectively.
As with the oxygen abundance gradients, there is no obvious correlation with these quantities.
Our isolated and quiescent galaxies (\ltgt and large $s_5$) show small positive [O/Fe]$_{\rm g}$ gradients, which means [O/Fe]$_{\rm g}$ is low at the centre.
This could be due to the ejecta of Type Ia supernovae remaining in the centre.

Finally, we note that our sample of galaxies with measurable gas-phase gradients is relatively small {, because we limit our sample to those with more than 1000 gas and young star particles (Section \ref{sec:fit}).
A simulation of larger volume and enough} resolution would be needed to better understand our gas-phase metallicity gradient results.

\section{Discussion and Conclusions}
\label{sec:5paper5_conc}

Our cosmological simulations, which include both AGN feedback and detailed chemical enrichment, allow us to predict the radial gradients of metallicity and elemental abundance ratios for both stars and gas.
 {The range used for determining gradients was chosen on a galaxy-by-galaxy basis in order to mask satellite galaxies and other noisy features in the metallicity profiles (see Section \ref{sec:fit}).}
We have presented these gradients both for the full population of galaxies at $z=0$, depending on the stellar populations ($\Delta$SFMS), the recent merging histories ($t_{\rm merge}$), and the environment ($s_5$).
 {We used $\Delta$SFMS and $t_{\rm merge}$ as proxies of morphological type, which cannot be fully resolved in our low-mass simulated galaxies.}
We have also shown the stellar metallicity gradient for the lifetime of two example galaxies that are in distinct environments.
The evolutionary paths of metallicity gradients are:
i) Assembly of gas-rich subgalaxies leads a high star formation rate and a steep initial metallicity gradient of stars ($z\gtsim2$).
ii) The gradients gradually become shallower due to star formation and/or stellar accretion at outskirts (passive evolution, or inside-out growth, $z\gtsim1$).
 {
iii) Major mergers cause stellar metallicity gradients to become shallower by redistributing high-metallicity stars from the centre to the outskirts.
iv) Central star formation induced by gas-rich mergers can regenerate metallicity gradients as well as creating steeper positive [O/Fe]$_*$ gradients.
v) However, galaxies with massive BHs and strong AGN feedback cannot regenerate their gradients, and are more susceptible to the effects of mergers.
}

Furthermore, we found the distance of a galaxy from the SFMS to be a good indicator of galaxy type in the simulations.
Those galaxies defined as \etgs had gradients that agree very well with the observational data for ETGs from the SAURON and ATLAS$^{\rm 3D}$ surveys.
We also find a turnover at $\sim 3\times10^{10}\msun$ as was found by \citet{spolaor09,spolaor10}.
The origin of this turnover can be understood as follows: at $\gtsim 3\times10^{10}\msun$, more massive galaxies tend to  {host stronger AGN and experience more mergers \citep[see][for more details]{pt17}.
Once AGN feedback has quenched star formation in these galaxies, the gradients can be flattened by mergers and not regenerate.
At $\ltsim 3\times10^{10}\msun$,}  {galaxies have shallower potentials, and can be more easily affected by supernova feedback to shut off star formation (Section \ref{sec:5paper5_starz0}).}

By comparing the distribution of metallicity gradients at $z=0$ obtained with the two different simulations, we have statistically shown that  {AGN feedback mainly affects the stellar metallicity gradients of massive ETGs (Figs. \ref{fig:starzhist} and \ref{fig:starofe_hist}), via the mechanism described above.
Since AGN feedback does not affect stars directly, it does not cause the metallicity gradients in these galaxies to flatten, rather, it prevents them from becoming steeper once they have been flattened by mergers.
It has no effect in less-massive galaxies where the BH does not grow to be massive enough to influence star formation significantly.}
While we find negative metallicity gradients for most simulated galaxies, [O/Fe]$_*$ tends to show small positive gradients; the central parts are more enriched by Type Ia supernovae.
This depends on the AGN feedback, and stronger AGN feedback might be required if the observational data show no significant [O/Fe]$_*$ gradients for all galaxies.

 {There is some AGN effect seen for the gas-phase [O/H] and [O/Fe] gradients, but this is secondary; massive galaxies are excluded from the sample with AGN because they do not have enough star formation at present.
For the galaxies that do not host strong AGN, we found no significant difference when AGN feedback was included or not.}
The oxygen abundance gradients obtained from the simulations showed a larger spread than observed values.
This could not be explained by either the merging history or environment of the galaxies,  {and might be due to the finite resolution of our simulations}.
In any case, simulations of larger volume and higher resolution will be required to fully investigate the driving factors behind gas-phase metallicity gradients.

\section*{Acknowledgements}
The authors are most grateful to H.~Kuntscher and R.~McDermid for providing early access to metallicity gradients from the ATLAS$^{\rm 3D}$ survey.
PT acknowledges funding from an STFC studentship.
This work was supported by a Discovery Projects grant from the Australian Research Council (grant DP150104329).
This work has made use of the University of Hertfordshire Science and Technology Research Institute high-performance computing facility.
This research was undertaken with the assistance of resources provided at the NCI National Facility systems at the Australian National University through the National Computational Merit Allocation Scheme supported by the Australian Government.
This research made use of the DiRAC HPC cluster at Durham. DiRAC is the UK HPC facility for particle physics, astrophysics, and cosmology, and is supported by STFC and BIS.
CK acknowledges PRACE for awarding her access to resource ARCHER based in the UK at Edinburgh.
Finally, we thank V. Springel for providing {\sc GADGET-3}.


\bibliographystyle{mn2e}
\bibliography{./refs}

\begin{thebibliography}{}

\bibitem[\protect\citeauthoryear{{Barnes}}{{Barnes}}{1988}]{barnes88}
{Barnes} J.~E.,  1988, \apj, 331, 699

\bibitem[\protect\citeauthoryear{{Bernardi}, {Shankar}, {Hyde}, {Mei},
  {Marulli} \& {Sheth}}{{Bernardi} et~al.}{2010}]{bernardi10}
{Bernardi} M.,  {Shankar} F.,  {Hyde} J.~B.,  {Mei} S.,  {Marulli} F.,
  {Sheth} R.~K.,  2010, \mnras, 404, 2087

\bibitem[\protect\citeauthoryear{{Bland-Hawthorn}}{{Bland-Hawthorn}}{2015}]{hector15}
{Bland-Hawthorn} J.,  2015, in {Ziegler} B.~L.,  {Combes} F.,  {Dannerbauer}
  H.,   {Verdugo} M.,  eds,  IAU Symposium Vol. 309, IAU Symposium. pp 21--28

\bibitem[\protect\citeauthoryear{{Brough}, {Tran}, {Sharp}, {von der Linden} \&
  {Couch}}{{Brough} et~al.}{2011}]{brough11}
{Brough} S.,  {Tran} K.-V.,  {Sharp} R.~G.,  {von der Linden} A.,    {Couch}
  W.~J.,  2011, \mnras, 414, L80

\bibitem[\protect\citeauthoryear{{Bundy} et~al.,}{{Bundy}
  et~al.}{2015}]{bundy15}
{Bundy} K.  et~al., 2015, \apj, 798, 7

\bibitem[\protect\citeauthoryear{{Cappellari} et~al.,}{{Cappellari}
  et~al.}{2011}]{cappellari11}
{Cappellari} M.  et~al., 2011, \mnras, 416, 1680

\bibitem[\protect\citeauthoryear{{Carlberg}}{{Carlberg}}{1984}]{carlberg84}
{Carlberg} R.~G.,  1984, \apj, 286, 403

\bibitem[\protect\citeauthoryear{{Cowie}, {Songaila}, {Hu} \& {Cohen}}{{Cowie}
  et~al.}{1996}]{cowie96}
{Cowie} L.~L.,  {Songaila} A.,  {Hu} E.~M.,    {Cohen} J.~G.,  1996, \aj, 112,
  839

\bibitem[\protect\citeauthoryear{{Cresci}, {Mannucci}, {Maiolino}, {Marconi},
  {Gnerucci} \& {Magrini}}{{Cresci} et~al.}{2010}]{cresci10}
{Cresci} G.,  {Mannucci} F.,  {Maiolino} R.,  {Marconi} A.,  {Gnerucci} A.,
  {Magrini} L.,  2010, \nat, 467, 811

\bibitem[\protect\citeauthoryear{{Croton} et~al.,}{{Croton}
  et~al.}{2006}]{croton06}
{Croton} D.~J.  et~al., 2006, \mnras, 365, 11

\bibitem[\protect\citeauthoryear{{Davies}, {Sadler} \& {Peletier}}{{Davies}
  et~al.}{1993}]{davies93}
{Davies} R.~L.,  {Sadler} E.~M.,    {Peletier} R.~F.,  1993, \mnras, 262, 650

\bibitem[\protect\citeauthoryear{{Dressler}}{{Dressler}}{1980}]{dressler80}
{Dressler} A.,  1980, \apj, 236, 351

\bibitem[\protect\citeauthoryear{{Few}, {Gibson}, {Courty}, {Michel-Dansac},
  {Brook} \& {Stinson}}{{Few} et~al.}{2012}]{few12}
{Few} C.~G.,  {Gibson} B.~K.,  {Courty} S.,  {Michel-Dansac} L.,  {Brook}
  C.~B.,    {Stinson} G.~S.,  2012, \aap, 547, A63

\bibitem[\protect\citeauthoryear{{Haardt} \& {Madau}}{{Haardt} \&
  {Madau}}{1996}]{haardt96}
{Haardt} F.,  {Madau} P.,  1996, \apj, 461, 20

\bibitem[\protect\citeauthoryear{{Hinshaw} et~al.,}{{Hinshaw}
  et~al.}{2013}]{wmap9}
{Hinshaw} G.  et~al., 2013, \apjs, 208, 19

\bibitem[\protect\citeauthoryear{{Ho} et~al.,}{{Ho} et~al.}{2014}]{ho14}
{Ho} I.-T.  et~al., 2014, \mnras, 444, 3894

\bibitem[\protect\citeauthoryear{{Ho}, {Kudritzki}, {Kewley}, {Zahid},
  {Dopita}, {Bresolin} \& {Rupke}}{{Ho} et~al.}{2015}]{ho15}
{Ho} I.-T.,  {Kudritzki} R.-P.,  {Kewley} L.~J.,  {Zahid} H.~J.,  {Dopita}
  M.~A.,  {Bresolin} F.,    {Rupke} D.~S.~N.,  2015, \mnras, 448, 2030

\bibitem[\protect\citeauthoryear{{Hopkins}, {Cox}, {Dutta}, {Hernquist},
  {Kormendy} \& {Lauer}}{{Hopkins} et~al.}{2009}]{hopkins09}
{Hopkins} P.~F.,  {Cox} T.~J.,  {Dutta} S.~N.,  {Hernquist} L.,  {Kormendy} J.,
     {Lauer} T.~R.,  2009, \apjs, 181, 135

\bibitem[\protect\citeauthoryear{{Houghton}}{{Houghton}}{2015}]{houghton15}
{Houghton} R.~C.~W.,  2015, \mnras, 451, 3427

\bibitem[\protect\citeauthoryear{{Jones}, {Ellis}, {Jullo} \&
  {Richard}}{{Jones} et~al.}{2010}]{jones10}
{Jones} T.,  {Ellis} R.,  {Jullo} E.,    {Richard} J.,  2010, \apjl, 725, L176

\bibitem[\protect\citeauthoryear{{Kauffmann} et~al.,}{{Kauffmann}
  et~al.}{2003}]{kauffmann03}
{Kauffmann} G.  et~al., 2003, \mnras, 341, 33

\bibitem[\protect\citeauthoryear{{Kobayashi}}{{Kobayashi}}{2004}]{ck04}
{Kobayashi} C.,  2004, \mnras, 347, 740

\bibitem[\protect\citeauthoryear{{Kobayashi} \& {Arimoto}}{{Kobayashi} \&
  {Arimoto}}{1999}]{ck99}
{Kobayashi} C.,  {Arimoto} N.,  1999, \apj, 527, 573

\bibitem[\protect\citeauthoryear{{Kobayashi}, {Karakas} \& {Umeda}}{{Kobayashi}
  et~al.}{2011}]{ck11b}
{Kobayashi} C.,  {Karakas} A.~I.,    {Umeda} H.,  2011, \mnras, 414, 3231

\bibitem[\protect\citeauthoryear{{Kobayashi} \& {Nakasato}}{{Kobayashi} \&
  {Nakasato}}{2011}]{ck11a}
{Kobayashi} C.,  {Nakasato} N.,  2011, \apj, 729, 16

\bibitem[\protect\citeauthoryear{{Kobayashi} \& {Nomoto}}{{Kobayashi} \&
  {Nomoto}}{2009}]{ck09}
{Kobayashi} C.,  {Nomoto} K.,  2009, \apj, 707, 1466

\bibitem[\protect\citeauthoryear{{Kobayashi}, {Springel} \&
  {White}}{{Kobayashi} et~al.}{2007}]{ck07}
{Kobayashi} C.,  {Springel} V.,    {White} S.~D.~M.,  2007, \mnras, 376, 1465

\bibitem[\protect\citeauthoryear{{Kroupa}}{{Kroupa}}{2008}]{kroupa08}
{Kroupa} P.,  2008, in {Knapen} J.~H.,  {Mahoney} T.~J.,   {Vazdekis} A.,  eds,
   Astronomical Society of the Pacific Conference Series Vol. 390, Pathways
  Through an Eclectic Universe. p.~3

\bibitem[\protect\citeauthoryear{{Kudritzki}, {Ho}, {Schruba}, {Burkert},
  {Zahid}, {Bresolin} \& {Dima}}{{Kudritzki} et~al.}{2015}]{kudritzki15}
{Kudritzki} R.-P.,  {Ho} I.-T.,  {Schruba} A.,  {Burkert} A.,  {Zahid} H.~J.,
  {Bresolin} F.,    {Dima} G.~I.,  2015, \mnras, 450, 342

\bibitem[\protect\citeauthoryear{{Kuntschner} et~al.,}{{Kuntschner}
  et~al.}{2010}]{kuntschner10}
{Kuntschner} H.  et~al., 2010, \mnras, 408, 97

\bibitem[\protect\citeauthoryear{{Lagos}, {Davis}, {Lacey}, {Zwaan}, {Baugh},
  {Gonzalez-Perez} \& {Padilla}}{{Lagos} et~al.}{2014}]{lagos14}
{Lagos} C.~d.~P.,  {Davis} T.~A.,  {Lacey} C.~G.,  {Zwaan} M.~A.,  {Baugh}
  C.~M.,  {Gonzalez-Perez} V.,    {Padilla} N.~D.,  2014, \mnras, 443, 1002

\bibitem[\protect\citeauthoryear{{Larson}}{{Larson}}{1974}]{larson74}
{Larson} R.~B.,  1974, \mnras, 166, 585

\bibitem[\protect\citeauthoryear{{Loubser} \&
  {S{\'a}nchez-Bl{\'a}zquez}}{{Loubser} \&
  {S{\'a}nchez-Bl{\'a}zquez}}{2011}]{loubser11}
{Loubser} S.~I.,  {S{\'a}nchez-Bl{\'a}zquez} P.,  2011, \mnras, 415, 3013

\bibitem[\protect\citeauthoryear{{Maciel}, {Lago} \& {Costa}}{{Maciel}
  et~al.}{2006}]{maciel06}
{Maciel} W.~J.,  {Lago} L.~G.,    {Costa} R.~D.~D.,  2006, \aap, 453, 587

\bibitem[\protect\citeauthoryear{{McGaugh}}{{McGaugh}}{2005}]{mcgaugh05}
{McGaugh} S.~S.,  2005, \apj, 632, 859

\bibitem[\protect\citeauthoryear{{McGaugh} \& {Schombert}}{{McGaugh} \&
  {Schombert}}{2015}]{mcgaugh15}
{McGaugh} S.~S.,  {Schombert} J.~M.,  2015, \apj, 802, 18

\bibitem[\protect\citeauthoryear{{Mori}, {Yoshii}, {Tsujimoto} \&
  {Nomoto}}{{Mori} et~al.}{1997}]{mori97}
{Mori} M.,  {Yoshii} Y.,  {Tsujimoto} T.,    {Nomoto} K.,  1997, \apjl, 478,
  L21

\bibitem[\protect\citeauthoryear{{Murante}, {Monaco}, {Giovalli}, {Borgani} \&
  {Diaferio}}{{Murante} et~al.}{2010}]{murante10}
{Murante} G.,  {Monaco} P.,  {Giovalli} M.,  {Borgani} S.,    {Diaferio} A.,
  2010, \mnras, 405, 1491

\bibitem[\protect\citeauthoryear{{Nomoto}, {Kobayashi} \& {Tominaga}}{{Nomoto}
  et~al.}{2013}]{nomoto13}
{Nomoto} K.,  {Kobayashi} C.,    {Tominaga} N.,  2013, \araa, 51, 457

\bibitem[\protect\citeauthoryear{{Ogando}, {Maia}, {Chiappini}, {Pellegrini},
  {Schiavon} \& {da Costa}}{{Ogando} et~al.}{2005}]{ogando05}
{Ogando} R.~L.~C.,  {Maia} M.~A.~G.,  {Chiappini} C.,  {Pellegrini} P.~S.,
  {Schiavon} R.~P.,    {da Costa} L.~N.,  2005, \apjl, 632, L61

\bibitem[\protect\citeauthoryear{{Pipino}, {D'Ercole}, {Chiappini} \&
  {Matteucci}}{{Pipino} et~al.}{2010}]{pipino10}
{Pipino} A.,  {D'Ercole} A.,  {Chiappini} C.,    {Matteucci} F.,  2010, \mnras,
  407, 1347

\bibitem[\protect\citeauthoryear{{Renzini} \& {Peng}}{{Renzini} \&
  {Peng}}{2015}]{renzini15}
{Renzini} A.,  {Peng} Y.-j.,  2015, \apjl, 801, L29

\bibitem[\protect\citeauthoryear{{Rupke}, {Kewley} \& {Chien}}{{Rupke}
  et~al.}{2010}]{rupke10}
{Rupke} D.~S.~N.,  {Kewley} L.~J.,    {Chien} L.-H.,  2010, \apj, 723, 1255

\bibitem[\protect\citeauthoryear{{S{\'a}nchez} et~al.,}{{S{\'a}nchez}
  et~al.}{2012}]{califa12}
{S{\'a}nchez} S.~F.  et~al., 2012, \aap, 538, A8

\bibitem[\protect\citeauthoryear{{S{\'a}nchez-Bl{\'a}zquez}
  et~al.,}{{S{\'a}nchez-Bl{\'a}zquez} et~al.}{2014}]{sanchez14}
{S{\'a}nchez-Bl{\'a}zquez} P.  et~al., 2014, \aap, 570, A6

\bibitem[\protect\citeauthoryear{{Scannapieco} et~al.,}{{Scannapieco}
  et~al.}{2012}]{scannapieco12}
{Scannapieco} C.  et~al., 2012, \mnras, 423, 1726

\bibitem[\protect\citeauthoryear{{Schaye} et~al.,}{{Schaye}
  et~al.}{2015}]{schaye15}
{Schaye} J.  et~al., 2015, \mnras, 446, 521

\bibitem[\protect\citeauthoryear{{Schweizer} \& {Seitzer}}{{Schweizer} \&
  {Seitzer}}{1992}]{schweizer92}
{Schweizer} F.,  {Seitzer} P.,  1992, \aj, 104, 1039

\bibitem[\protect\citeauthoryear{{Sijacki}, {Vogelsberger}, {Genel},
  {Springel}, {Torrey}, {Snyder}, {Nelson} \& {Hernquist}}{{Sijacki}
  et~al.}{2015}]{sijacki15}
{Sijacki} D.,  {Vogelsberger} M.,  {Genel} S.,  {Springel} V.,  {Torrey} P.,
  {Snyder} G.~F.,  {Nelson} D.,    {Hernquist} L.,  2015, \mnras, 452, 575

\bibitem[\protect\citeauthoryear{{Spolaor}, {Kobayashi}, {Forbes}, {Couch} \&
  {Hau}}{{Spolaor} et~al.}{2010}]{spolaor10}
{Spolaor} M.,  {Kobayashi} C.,  {Forbes} D.~A.,  {Couch} W.~J.,    {Hau}
  G.~K.~T.,  2010, \mnras, 408, 272

\bibitem[\protect\citeauthoryear{{Spolaor}, {Proctor}, {Forbes} \&
  {Couch}}{{Spolaor} et~al.}{2009}]{spolaor09}
{Spolaor} M.,  {Proctor} R.~N.,  {Forbes} D.~A.,    {Couch} W.~J.,  2009,
  \apjl, 691, L138

\bibitem[\protect\citeauthoryear{{Springel}, {Di Matteo} \&
  {Hernquist}}{{Springel} et~al.}{2005}]{springel05}
{Springel} V.,  {Di Matteo} T.,    {Hernquist} L.,  2005, \mnras, 361, 776

\bibitem[\protect\citeauthoryear{{Springel}, {White}, {Tormen} \&
  {Kauffmann}}{{Springel} et~al.}{2001}]{springel01}
{Springel} V.,  {White} S.~D.~M.,  {Tormen} G.,    {Kauffmann} G.,  2001,
  \mnras, 328, 726

\bibitem[\protect\citeauthoryear{{Sutherland} \& {Dopita}}{{Sutherland} \&
  {Dopita}}{1993}]{sutherland93}
{Sutherland} R.~S.,  {Dopita} M.~A.,  1993, \apjs, 88, 253

\bibitem[\protect\citeauthoryear{{Taylor}, {Federrath} \& {Kobayashi}}{{Taylor}
  et~al.}{2017}]{pt17}
{Taylor} P.,  {Federrath} C.,    {Kobayashi} C.,  2017, \mnras, 469, 4249

\bibitem[\protect\citeauthoryear{{Taylor} \& {Kobayashi}}{{Taylor} \&
  {Kobayashi}}{2014}]{pt14}
{Taylor} P.,  {Kobayashi} C.,  2014, \mnras, 442, 2751

\bibitem[\protect\citeauthoryear{{Taylor} \& {Kobayashi}}{{Taylor} \&
  {Kobayashi}}{2015a}]{pt15a}
{Taylor} P.,  {Kobayashi} C.,  2015a, \mnras, 448, 1835

\bibitem[\protect\citeauthoryear{{Taylor} \& {Kobayashi}}{{Taylor} \&
  {Kobayashi}}{2015b}]{pt15b}
{Taylor} P.,  {Kobayashi} C.,  2015b, \mnras, 452, L59

\bibitem[\protect\citeauthoryear{{Taylor} \& {Kobayashi}}{{Taylor} \&
  {Kobayashi}}{2016}]{pt16}
{Taylor} P.,  {Kobayashi} C.,  2016, \mnras, 463, 2465

\bibitem[\protect\citeauthoryear{{Tempel}, {Saar}, {Liivam{\"a}gi}, {Tamm},
  {Einasto}, {Einasto} \& {M{\"u}ller}}{{Tempel} et~al.}{2011}]{tempel11}
{Tempel} E.,  {Saar} E.,  {Liivam{\"a}gi} L.~J.,  {Tamm} A.,  {Einasto} J.,
  {Einasto} M.,    {M{\"u}ller} V.,  2011, \aap, 529, A53

\bibitem[\protect\citeauthoryear{{Tissera}, {Machado}, {Sanchez-Blazquez},
  {Pedrosa}, {S{\'a}nchez}, {Snaith} \& {Vilchez}}{{Tissera}
  et~al.}{2016}]{tissera16}
{Tissera} P.~B.,  {Machado} R.~E.~G.,  {Sanchez-Blazquez} P.,  {Pedrosa} S.~E.,
   {S{\'a}nchez} S.~F.,  {Snaith} O.,    {Vilchez} J.,  2016, \aap, 592, A93

\bibitem[\protect\citeauthoryear{{Toomre} \& {Toomre}}{{Toomre} \&
  {Toomre}}{1972}]{toomre72}
{Toomre} A.,  {Toomre} J.,  1972, \apj, 178, 623

\bibitem[\protect\citeauthoryear{{White}}{{White}}{1980}]{white80}
{White} S.~D.~M.,  1980, \mnras, 191, 1P

\bibitem[\protect\citeauthoryear{{Wuyts} et~al.,}{{Wuyts}
  et~al.}{2011}]{wuyts11}
{Wuyts} S.  et~al., 2011, \apj, 742, 96

\bibitem[\protect\citeauthoryear{{Young} et~al.,}{{Young}
  et~al.}{2011}]{young11}
{Young} L.~M.  et~al., 2011, \mnras, 414, 940

\bibitem[\protect\citeauthoryear{{Yuan}, {Kewley}, {Swinbank}, {Richard} \&
  {Livermore}}{{Yuan} et~al.}{2011}]{yuan11}
{Yuan} T.-T.,  {Kewley} L.~J.,  {Swinbank} A.~M.,  {Richard} J.,    {Livermore}
  R.~C.,  2011, \apjl, 732, L14

\bibitem[\protect\citeauthoryear{{Zabludoff}, {Zaritsky}, {Lin}, {Tucker},
  {Hashimoto}, {Shectman}, {Oemler} \& {Kirshner}}{{Zabludoff}
  et~al.}{1996}]{zabludoff96}
{Zabludoff} A.~I.,  {Zaritsky} D.,  {Lin} H.,  {Tucker} D.,  {Hashimoto} Y.,
  {Shectman} S.~A.,  {Oemler} A.,    {Kirshner} R.~P.,  1996, \apj, 466, 104

\bibitem[\protect\citeauthoryear{{Zaritsky}, {Kennicutt} Jr. \&
  {Huchra}}{{Zaritsky} et~al.}{1994}]{zaritsky94}
{Zaritsky} D.,  {Kennicutt} Jr. R.~C.,    {Huchra} J.~P.,  1994, \apj, 420, 87

\end{thebibliography}

\appendix

 {
\section{Linear fits for Stellar Metallicity Gradients}
\label{ap:sanchez}
}
\begin{figure}
	\centering
	\includegraphics[width=0.48\textwidth,keepaspectratio]{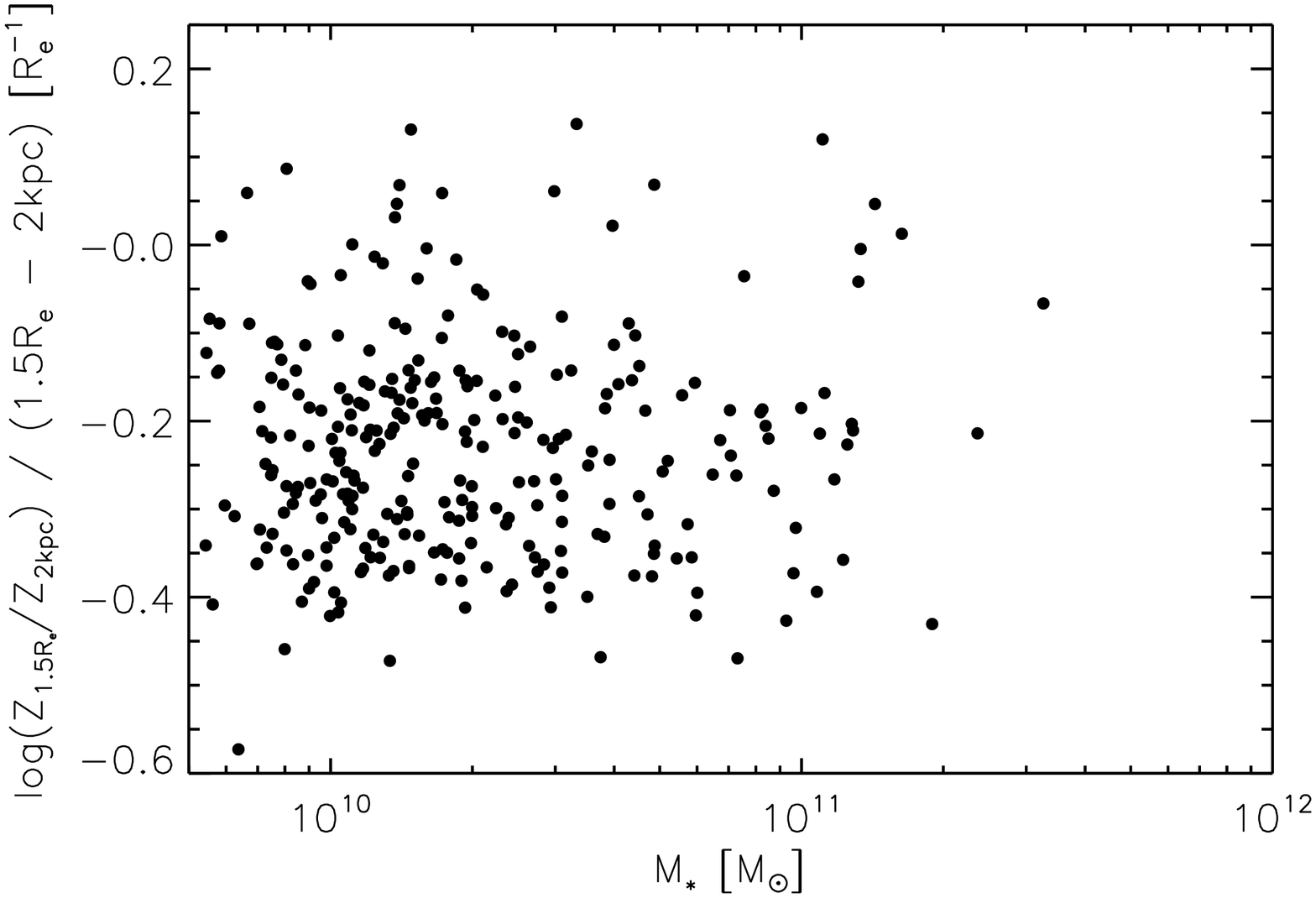}
	\caption{Stellar metallicity gradient in units of dex $R_{\rm e}^{-1}$ for simulated galaxies defined as \ltgs as a function of mass.}
	\label{fig:starzlin}
\end{figure}

 {\citet{sanchez14} studied the stellar metallicity gradients in 62 face-on spiral galaxies from the CALIFA survey.
They derive metallicity gradients by two methods: fitting a linear profile of the form $\log Z = a + br$ to the disc of the galaxy; or calculating the difference in metallicity at $1.5 R_{\rm e}$ and the bulge radius.
Both methods are shown to give similar results, and the second is adopted throughout the paper.
To compare to this result, in this section, we define the quantity $\nabla Z_{\rm linear}$ as 
\begin{equation}
	\nabla Z_{\rm linear} = \frac{\log\left( Z_{1.5 R_{\rm e}}/Z_\odot\right) - \log\left(Z_{2\,{\rm kpc}}/Z_\odot\right)}{1.5 R_{\rm e} - 2\,{\rm kpc}}.
\end{equation}
Although the bulge and disc components are not fully resolved in our cosmological simulations, the inner radius of 2 kpc is large enough to exclude the bulge components in metallicity gradients for zoom-in disc galaxy simulations \citep[e.g.,][]{ck11a}.}

 {
$\nabla Z_{\rm linear}$ is calculated for all galaxies defined as \ltgs in the simulation with AGN feedback, and the results are shown in Fig. \ref{fig:starzlin}.
Consistent with \citet{sanchez14}, there is no trend with galaxy mass, and the $\nabla Z_{\rm linear}$ values occupy a similar range.
However, we note that many of the metallicity profiles for these galaxies would not have been well fit by a straight line.
}

\section{Merger Trees}
\label{ap:tree}

\begin{figure*}
	\centering
	\includegraphics[width=\textwidth,keepaspectratio]{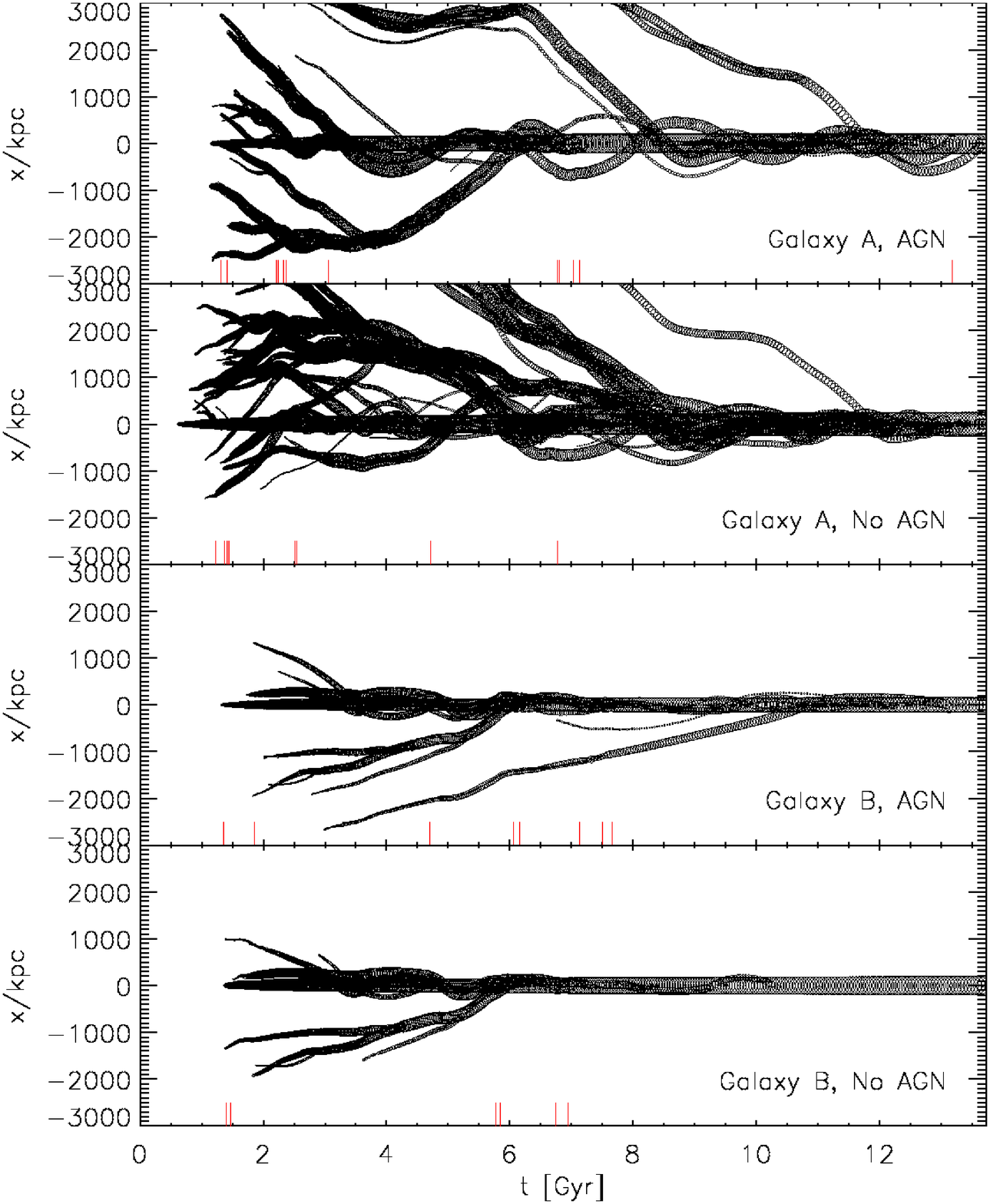}
	\caption{Merger trees for galaxies A and B (top two and bottom two panels, respectively) with and without AGN feedback (first \& third, and second \& fourth panels, respectively).
	The radius of the plotted circles is proportional to the logarithm of stellar mass, and one component of the separation vector from the galaxy of focus is shown.
	Vertical red lines indicate when the galaxy is identified as experiencing a major merger.}
	\label{fig:trees}
\end{figure*}

We identify galaxies with a Friends-of-Friends algorithm for all particle types, and construct merger trees by matching particles by ID number between snapshots.
Fig. \ref{fig:trees} shows merger trees for galaxies A and B (top two and bottom two panels, respectively) as a function of cosmic time, for the simulations with and without AGN feedback (first \& third, and second \& fourth panels, respectively).
In each panel, one component of separation from the galaxy in question is shown, and the size of plotted circles is proportional to the logarithm of galaxy stellar mass.
The vertical red lines in each panel show when the galaxy is identified as undergoing a major merger (stellar mass ratio greater than 1:4).
The last epoch of major merger defines $t_{\rm merge}$.

\section{$n^{\rm th}$ Nearest Neighbour Parameters}
\label{ap:s5}
We describe how the parameters of the $n^{\rm th}$ nearest neighbour calculation affect out results, namely the value of $n$ and galaxy mass cut used, illustrated using the $z=0$ relation between stellar metallicity gradient and stellar mass.
We showed in Fig. \ref{fig:s5} (Section \ref{sec:morph}) the distribution of galaxies in one of our simulations, coloured by their $5^{\rm th}$ nearest neighbour distance, $s_5$.
As expected, those galaxies in the densest environments have the smallest $s_5$.
Note that $s_5$ is calculated in 3-D, not in projection.

\begin{figure}
	\centering
	\includegraphics[width=0.48\textwidth,keepaspectratio]{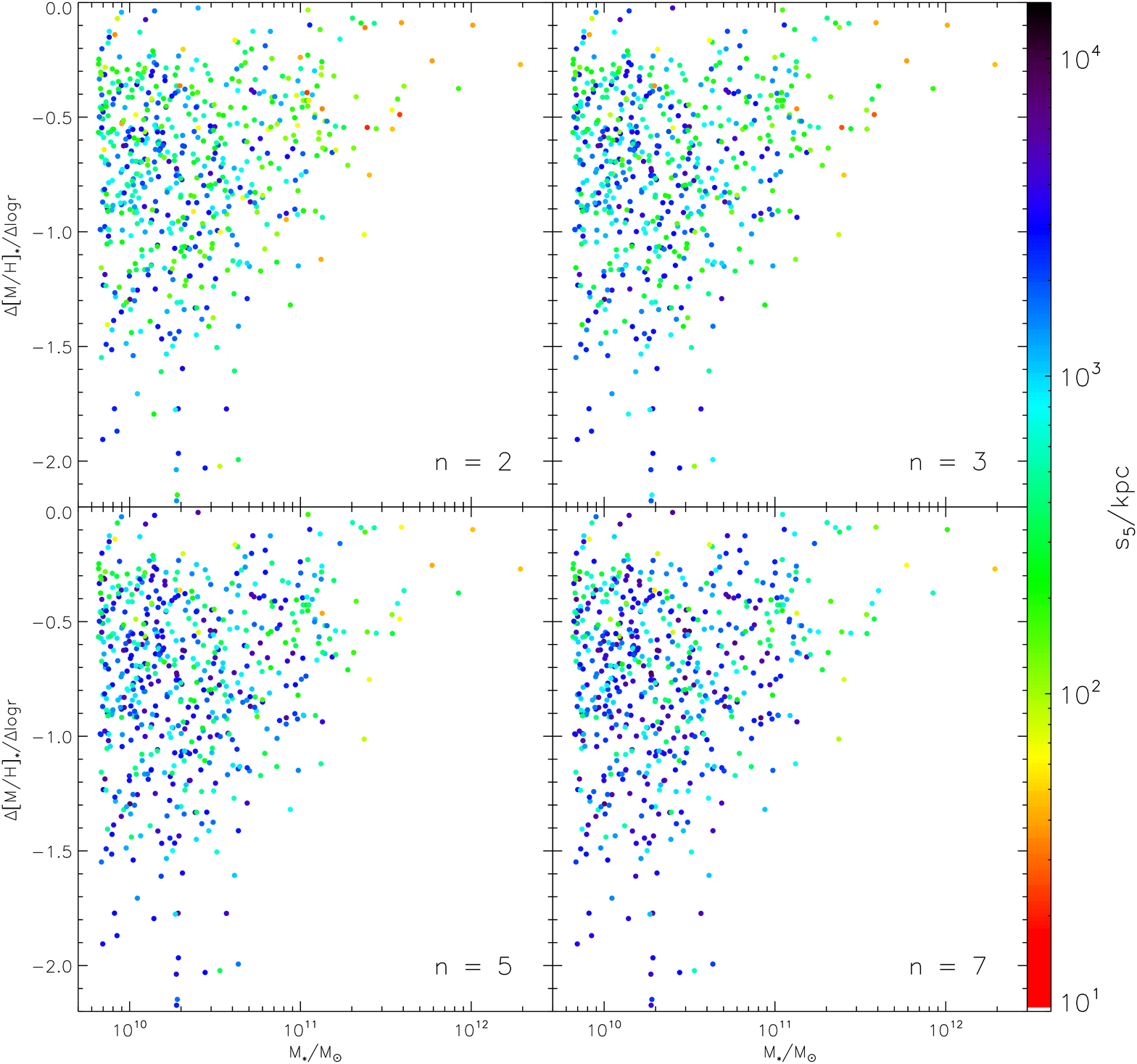}
	\caption{The relation between stellar metallicity gradient and stellar mass at $z=0$ from our simulations, coloured by $n^{\rm th}$ nearest neighbour distance, $s_n$.
	From top left to bottom right, we use $n=2,\,3,\,5,\,7$.
	In all panels, all galaxies are used in the calculation.}
	\label{fig:appn}
\end{figure}

Fig. \ref{fig:appn} shows that, as one would expect, as $n$ increases, so too does $s_n$, and that the trend seen in the data, whereby galaxies with shallower gradients are more likely to have with smaller $s_n$, is clear for all values of $n$ used.
The value of $n$ used here is less important than in observational works where an estimate of the galaxy density is often required ($\sim1/\pi s_n^2$), whereas we are not trying to quantify the effects of environment on galaxy evolution.

\begin{figure}
	\centering
	\includegraphics[width=0.48\textwidth,keepaspectratio]{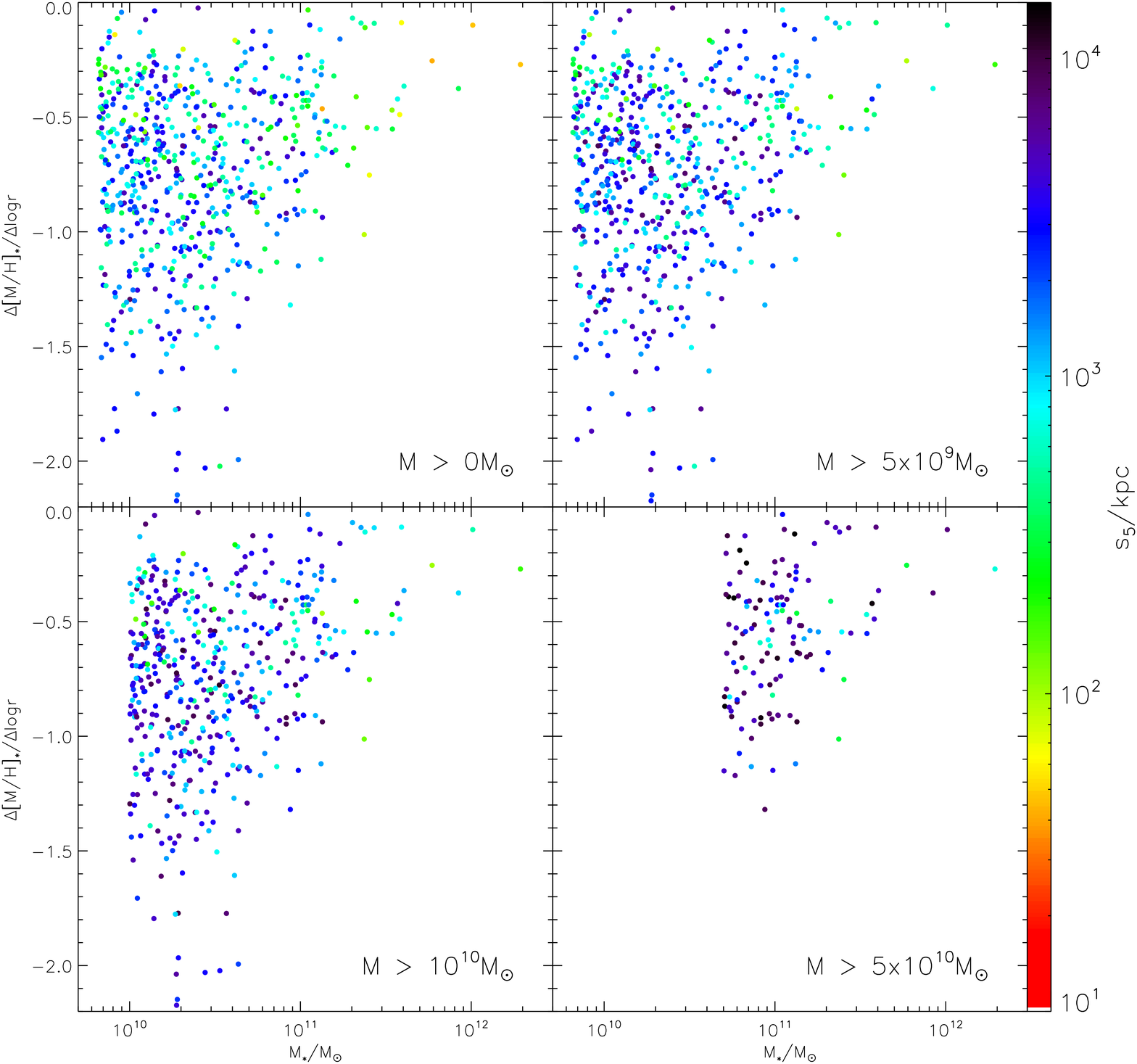}
	\caption{The relation between stellar metallicity gradient and stellar mass at $z=0$ from our simulations, coloured by $5^{\rm th}$ nearest neighbour distance.
	Only galaxies above a given mass limit are considered in each panel; from top left to bottom right these limits are $0,\,5\times10^9,\,10^{10},\,5\times10^{10}M_\odot$.}
	\label{fig:appm}
\end{figure}

Fig. \ref{fig:appm} shows how our results change if we impose a lower mass limit, as is often the case in observations.
As the mass limit increases, the identified trend becomes harder to see, and with $M_*>10^{10}\msun$ is no longer present.
This happens because the most massive galaxies are distributed on scales comparable to the size of the simulation box, and certainly on scales much larger than environmental effects influence galaxy evolution.


\bsp

\label{lastpage}

\end{document}